\documentclass{article}

\usepackage{epsf, hyperref, graphicx, amssymb, amsmath, multirow, enumitem, tabto, textcomp, gensymb}
\hypersetup{final}

\newif\ifCS
\CSfalse
\ifCS
    \usepackage{ComplexSystems}
    \copyrightinfo{Volume}{year}{1}{1+}
\else
\fi

\begin{document}

\title{Lenia --- Biology of Artificial Life}

\ifCS
    \author{\authname{Bert Wang-Chak Chan}\\[2pt] 
    \authadd{Hong Kong}}
    \markboth{Complex Systems} 
    {Lenia --- Biology of Artificial Life} 
\else
    \author{Bert Wang-Chak Chan\thanks{albert.chak@gmail.com}\\ \textit{Hong Kong}}
    \date{Mar 2019}
\fi

\maketitle

\begin{abstract}
We report a new system of artificial life called \textit{Lenia} (from Latin \mbox{\textit{lenis}} ``smooth''), a two-dimensional cellular automaton with continuous space-time-state and generalized local rule.  Computer simulations show that Lenia supports a great diversity of complex autonomous patterns or ``lifeforms'' bearing resemblance to real-world microscopic organisms.  More than 400 species in 18 families have been identified, many discovered via interactive evolutionary computation.  They differ from other cellular automata patterns in being geometric, metameric, fuzzy, resilient, adaptive, and rule-generic.

We present basic observations of the system regarding the properties of space-time and basic settings.  We provide a broad survey of the lifeforms, categorize them into a hierarchical taxonomy, and map their distribution in the parameter hyperspace.  We describe their morphological structures and behavioral dynamics, propose possible mechanisms of their self-propulsion, self-organization and plasticity.  Finally, we discuss how the study of Lenia would be related to biology, artificial life, and artificial intelligence.
\end{abstract}

\ifCS
\else
\fi

\section{Introduction}

Among the long-term goals of \textit{artificial life} are to simulate existing biological life and to create new life forms using artificial systems.  These are expressed in the fourteen open problems in artificial life \cite{Bedau2000}, in which number three is of particular interest here:
\begin{quote}
\textit{Determine whether fundamentally novel living organizations can \mbox{exist}.}
\end{quote}

There have been numerous efforts in creating and studying novel mathematical systems that are capable of simulating complex life-like dynamics.  Examples include particle systems like Swarm Chemistry \cite{Sayama2009}, Primordial Particle Systems (PPS) \cite{Schmickl2016}; reaction-diffusion systems like the U-Skate World \cite{Munafo2014}; cellular automata like the Game of Life (GoL) \cite{Gardner1970}, elementary cellular automata (ECA) \cite{Wolfram1983}; evolutionary systems like virtual creatures \cite{Sims1994}, soft robots \cite{Cheney2013, Kriegman2018}.  These systems have a common theme --- let there be countless modules or particles and (often localized) interactions among them, a complex system with interesting autonomous patterns will emerge, just like how life emerged on Earth 4.28 billion years ago \cite{Dodd2017}.

Life can be defined as the capabilities of self-organizing (morphogenesis), self-regulating (homeostasis), self-directing (motility), self-replicating (reproduction), entropy reduction (metabolism), growth (development), response to stimuli (sensitivity), response to environment (adaptability), and evolving through mutation and selection (evolvability) (e.g., \cite{Koshland2002, McKay2004, Sagan1970, Schrodinger1967}).  Systems of artificial life are able to reproduce some of these capabilities with various levels of fidelity.  Lenia, the subject of this paper, is able to achieve many of them, except self-replication that is yet to be discovered.

Lenia also captures a few subjective characteristics of life, like vividness, fuzziness, aesthetic appeal, and the great diversity and subtle variety in patterns that a biologist would have the urge to collect and catalogue them.  If there is some truth in the biophilia hypothesis \cite{Wilson1984} that humans are innately attracted to nature, it may not be too far-fetched to suggest that these subjective experiences are not merely feelings but among the essences of life as we know it.

Due to similarities between life on Earth and Lenia, we borrow terminologies and concepts from biology, like taxonomy (corresponds to categorization), binomial nomenclature (naming), ecology (parameter space), morphology (structures), behavior (dynamics), physiology (mechanisms), and allometry (statistics).  We also borrow space-time (grid and time-step) and fundamental laws (local rule) from physics.  With a few caveats, these borrowings are useful in providing more intuitive characterization of the system, and may facilitate discussions on how Lenia or similar systems could give answers to life \cite{Langton1986}, the universe \cite{Wolfram2002}, and everything.

\subsection{Background}

A \textit{cellular automaton} (CA, plural: cellular automata) is a mathematical system where a grid of sites, each having a particular state at a moment, are being updated repeatedly according to a local rule and each site's neighboring sites.  Since its conception by John von Neumann and Stanislaw Ulam \cite{VonNeumann1951, Ulam1962}, various CAs have been investigated, the most famous being Stephen Wolfram's one-dimensional elementary cellular automata (ECA) \cite{Wolfram1983, Wolfram2002} and John H. Conway's two-dimensional \textit{Game of Life} (GoL) \cite{Gardner1970, Adamatzky2010}.  GoL is the starting point of where Lenia came from.  It produces a whole universe of interesting patterns \cite{LifeWiki} ranging from simple ``still lifes'', ``oscillators'' and ``spaceships'', to complex constructs like pattern emitters, self-replicators, and even fully operational computers thanks to its Turing completeness \cite{Rendell2002}.

\begin{table}[!tb]
\centerline{\scriptsize\begin{tabular}{|ll|cccc|}
\hline
\# &System &Type* &Space &Neighborhood &N. sum \\
\hline
1 &ECA, GoL &CA &singular &nearest cube &totalistic \\
2 &Continuous ECA &CA &singular &nearest cube &totalistic \\
3 &Continuous GoL &EA &continuous &continuous ball &totalistic \\
4 &Primordia &CA &singular &nearest cube &totalistic \\
\hline
5 &Larger-than-Life &CA / GCA &fractional &extended cube &totalistic \\
6 &RealLife &EA &continuous &continuous cube &totalistic \\
7 &SmoothLife &GCA &fractional &extended shell &totalistic \\
8 &Discrete Lenia &GCA &fractional &extended ball &weighted \\
9 &Continuous Lenia &EA &continuous &continuous ball &weighted \\
\hline
\hline
\# &Ref. and notes &Growth &Update &Time &State \\
\hline
1 &\cite{Wolfram2002, Gardner1970} &intervals &replace &singular &singular \\
2 &\cite{Wolfram2002} &mapping &replace &singular &continuous \\
3 &\cite{MacLennan1990} &mapping &replace &singular &continuous \\
4 &$\dagger$ &intervals &replace &singular &extended \\
\hline
5 &\cite{Griffeath1994, Evans2001} &intervals &replace &singular &singular \\
6 &\cite{Pivato2007} &intervals &replace &singular &singular \\
7 &\cite{Rafler2011}$\ddagger$ &intervals &increment &fractional &fractional \\
8 & &mapping &increment &fractional &fractional \\
9 & &mapping &increment &continuous &continuous \\
\hline
\end{tabular}}
\caption{Comparison of genericity and continuity in various CAs.  {\footnotesize (*~GCA = geometrical cellular automata, EA = Euclidean automata, see ``Discussion''.  $\dagger$~Primordia is a precursor to Lenia, written in JavaScript/HTML by the author circa 2005.  It had multi-states and extended survival/birth intervals.  $\ddagger$~SmoothLife and Lenia, being independent developments, exhibit striking resemblance in system and generated patterns.  This can be considered an instance of ``convergent evolution''.)}}
\label{table-ca}
\end{table}

\begin{figure}[!tb]
\centering
\includegraphics[width=\textwidth,trim=4 4 4 4,clip]{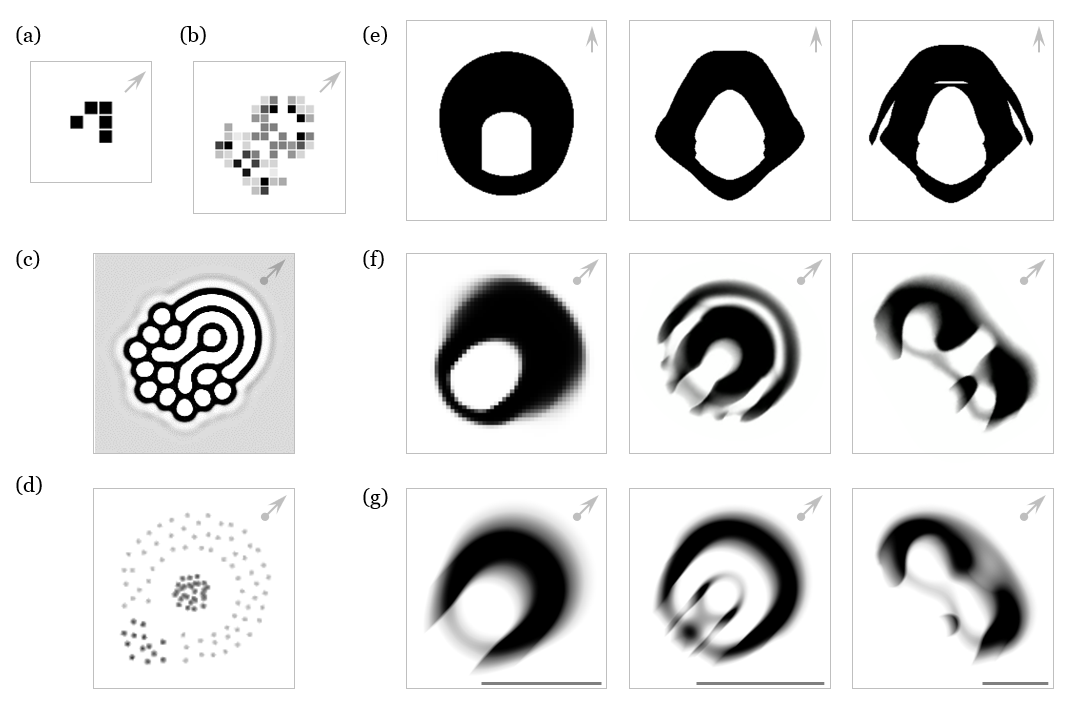}
\caption{Patterns in artificial life systems: cellular automata (a-b, e-g), reaction-diffusion (b) and particle swarm (c).  ($\uparrow$= orthogonal; $\nearrow$= diagonal; $\bullet\!\!^\nearrow$= omnidirectional; scale bar is unit length = kernel radius).  \textbf{(a)} Game of Life (GoL): ``glider''.  \textbf{(b)} Primordia: ``DX:8/762''.  \textbf{(c)} U-Skate World: ``Jellyfish'' \cite{Munafo2009}.  \textbf{(d)} Swarm Chemistry: ``Fast Walker \& Slow Follower'' \cite{Sayama2018b}.  \textbf{(e)} Larger-than-Life (LtL): ``bug with stomach'' using ball neighborhood, ``bug with ribbed stomach'', ``bug with wings'' \cite{Evans2003}.  \textbf{(f)} SmoothLife: ``smooth glider'', ``pulsating glider'', ``wobbly glider'' \cite{Rafler2011, Hutton2012a, Berger2017}.  \textbf{(g)} Lenia: \textit{Scutium}, \textit{Kronium}, \textit{Pyroscutium}.}
\label{fig-alife}
\end{figure}

Several aspects of GoL can be generalized.  A discrete \textit{singular} property (e.g. dead-or-alive state) can be \textit{extended} into a range (multi-state), normalized to a \textit{fractional} property in the unit range, and becomes \textit{continuous} by further splitting the range into infinitesimals (real number state).  The local rule can be generalized from the basic ECA/GoL style (e.g. totalistic sum) to smooth parameterized operations (weighted sum).

By comparing various CAs that possess autonomous soliton patterns, we observe the \textit{evolution of generalization} with increasing genericity and continuity (Table \ref{table-ca}, Figure \ref{fig-alife}).  This suggests that Lenia is currently at the latest stage of generalizing GoL, although there may be room for further generalizations.

\section{METHODS}

We describe the methods of constructing and studying Lenia, including its mathematical definition, computer simulation, strategies of evolving new lifeforms, and how to perform observational and statistical analysis.

\subsection{Definitions}

Mathematically, a CA is defined by a 5-tuple\footnote{Conventionally a 4-tuple with the timeline $\mathcal{T}$ omitted.} $\mathcal{A} = (\mathcal{L}, \mathcal{T}, \mathcal{S}, \mathcal{N}, \phi)$, where $\mathcal{L}$ is the $d$-dimensional \textit{lattice} or \textit{grid}, $\mathcal{T}$ is the \textit{timeline}, $\mathcal{S}$ is the \textit{state set}, $\mathcal{N} \subset \mathcal{L}$ is the \textit{neighborhood} of the origin, $\phi: \mathcal{S^N} \rightarrow \mathcal{S}$ is the \textit{local rule}.

Define $\mathbf{A}^t: \mathcal{L} \rightarrow \mathcal{S}$ as a \textit{configuration} or \textit{pattern} (i.e. collection of states over the whole grid) at time $t \in \mathcal{T}$. $\mathbf{A}^t(\mathbf{x})$ is the state of site $\mathbf{x} \in \mathcal{L}$, and $\mathbf{A}^t(\mathcal{N}_\mathbf{x}) = \{\mathbf{A}^t(\mathbf{n}): \mathbf{n} \in \mathcal{N}_\mathbf{x}\}$ is the state collection over the site’s neighborhood $\mathcal{N}_\mathbf{x} = \{\mathbf{x} + \mathbf{n} : \mathbf{n} \in \mathcal{N}\}$.  The \textit{global rule} is $\Phi: \mathcal{S^L} \rightarrow \mathcal{S^L}$ such that $\Phi(\mathbf{A})(\mathbf{x}) = \phi(\mathbf{A}\left(\mathcal{N}_\mathbf{x}) \right)$.
Starting from an initial configuration $\mathbf{A}^0$, the grid is updated according to the global rule $\Phi$ for each time-step $\Delta t$, leading to the following time-evolution:
\begin{equation}
\Phi(\mathbf{A}^0) = \mathbf{A}^{\Delta t}, \Phi(\mathbf{A}^{\Delta t}) = \mathbf{A}^{2\Delta t}, \ldots, \Phi(\mathbf{A}^t) = \mathbf{A}^{t+\Delta t}, \ldots
\end{equation}
After $N$ repeated updates (or \textit{generations}):
\begin{equation}
\Phi^N(\mathbf{A}^t) = \mathbf{A}^{t+N\Delta t}
\end{equation}

\begin{figure}[!tb]
\centering
\includegraphics[width=\textwidth,trim=4 4 4 4,clip]{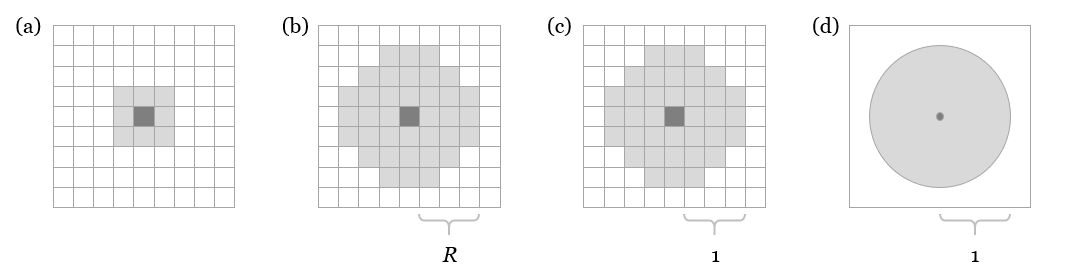}
\caption{Neighborhoods in various CAs.  \textbf{(a)} 8-site Moore neighborhood in GoL.  \textbf{(b-d)} Neighborhoods in Lenia, including range R extended neighborhood (b) and its normalization (c) in discrete Lenia, and the unit ball neighborhood in continuous Lenia (d).}
\label{fig-neigh}
\end{figure}

\begin{figure}[!tb]
\centering
\includegraphics[width=\textwidth,trim=4 4 4 4,clip]{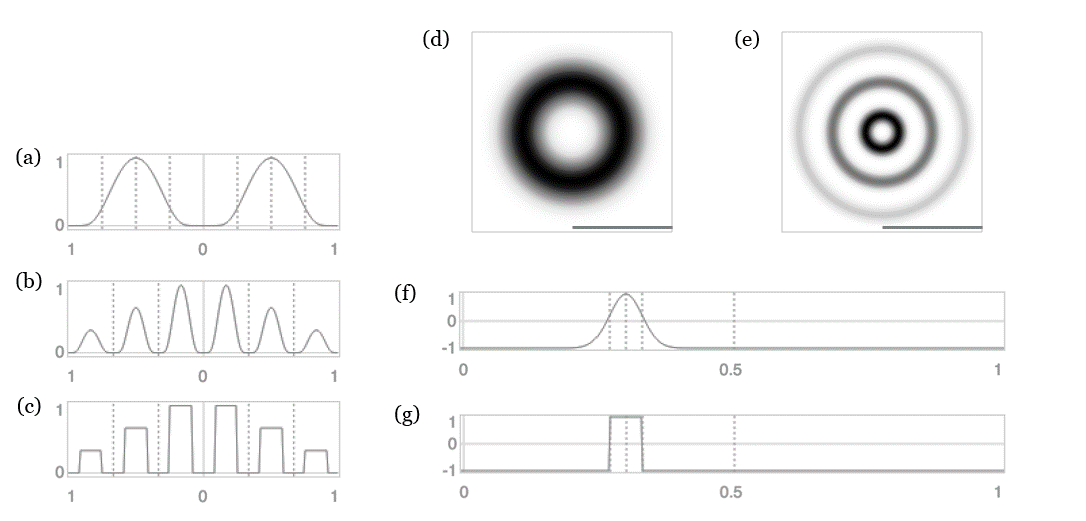}
\caption{Core functions in Lenia.  \textbf{(a-c)} Cross-section of the kernel: kernel core $K_C(r)$ using exponential function (a), and kernel shell $K_S(r; \beta)$ with peaks $\beta = (1, \frac{2}{3}, \frac{1}{3})$ using exponential (b) or rectangular core function (c).  \textbf{(d-e)} Kernel core (d) and kernel shell (e) as displayed in the grid, showing the ``influence'' (convolution weight) of the site on its neighborhood (darker = larger weight, more influence).  \textbf{(f-g)} Growth mapping $G(u; \mu, \sigma)$ with $\mu=0.3, \sigma=0.03$ using exponential (f) or rectangular (g) function.}
\label{fig-core}
\end{figure}

\subsubsection{Definition of Game of Life}
Take GoL as an example, $\mathcal{A}_\text{GoL} = (\mathcal{L}, \mathcal{T}, \mathcal{S}, \mathcal{N}, \phi)$, where $\mathcal{L} = \mathbb{Z}^2$ is the two-dimensional discrete grid; $\mathcal{T} = \mathbb{Z}$ is the discrete timeline; $\mathcal{S} = \{0, 1\}$ is the singular state set; $\mathcal{N} = \{-1, 0, 1\}^2$ is the Moore neighborhood (Chebyshev $L^\infty$ norm) including the site itself and its 8 neighbors (Figure \ref{fig-neigh}(a)).

The totalistic neighborhood sum of site $\mathbf{x}$ is:
\begin{equation}
\mathbf{S}^t(\mathbf{x}) = \sum_{n\in\mathcal{N}} \mathbf{A}^t(\mathbf{x}+\mathbf{n})
\end{equation}
Every site is updated synchronously according to the local rule:
\begin{equation}
\mathbf{A}^{t+1}(\mathbf{x}) =
\begin{cases}
1 & \text{ if } \mathbf{A}^t(\mathbf{x}) = 0 \text{ and } \mathbf{S}^t(\mathbf{x}) \in \{3\} \text{ (birth)} \\
1 & \text{ if } \mathbf{A}^t(\mathbf{x}) = 1 \text{ and } \mathbf{S}^t(\mathbf{x}) \in \{3, 4\} \text{ (survival)} \\
0 & \text{ otherwise} \text{ (death)}
\end{cases}
\end{equation}

\subsubsection{Definition of Lenia}
\textit{Discrete Lenia} (DL) generalizes GoL by extending and normalizing the space-time-state dimensions.  DL is used for computer simulation and analysis, and with normalization, patterns from different dimensions can be compared.

The state set is extended to $\mathcal{S} = \{0, 1, 2, \ldots, P\}$ with maximum $P \in \mathbb{Z}$.  The neighborhood is extended to a discrete ball (Euclidean $L^2$ norm) $\mathcal{N} = \mathbb{B}_R[0] =  \{\mathbf{x} \in \mathcal{L} : \|\mathbf{x}\|_2 \leq R\}$ of range $R \in \mathbb{Z}$ (Figure \ref{fig-neigh}(b)).

To normalize, define or redefine $R, T, P \in \mathbb{Z}$ as the \textit{space}, \textit{time} and \textit{state resolutions}, and their reciprocals $\Delta x = 1/R, \Delta t = 1/T, \Delta p = 1/P$ as the \textit{site distance}, \textit{time step}, and \textit{state precision}, respectively.  The dimensions are scaled by the reciprocals (Figure \ref{fig-neigh}(c)):
\begin{equation}
\mathcal{L} = \Delta x \; \mathbb{Z}^2, \ \mathcal{T} = \Delta t \; \mathbb{Z}, \ \mathcal{S} = \Delta p \; \{0 \ldots P\}, \ \mathcal{N} = \mathbb{B}_1 [0]
\end{equation}

\textit{Continuous Lenia} (CL) is hypothesized to exist as the resolutions of DL approach infinity $R , T, P \rightarrow \infty$ and the steps $\Delta x, \Delta t, \Delta p$ become infinitesimals $\mathrm{d}x, \mathrm{d}t, \mathrm{d}p$, the dimensions will approach their continuum limits, i.e. the Euclidean space, the real timeline, the unit interval states, and the continuous unit ball neighborhood (Figure \ref{fig-neigh}(d)):
\begin{equation}
\mathcal{L} = \mathbb{R}^2, \ \mathcal{T} = \mathbb{R}, \ \mathcal{S} = [0, 1], \ \mathcal{N} = \mathbb{B}_1 [0]
\end{equation}

However, there is a cardinality leap between the discrete dimensions in DL and the continuous dimensions in CL.  The existence of the continuum limit for space was proved mathematically in \cite{Pivato2007}, and our computer simulations provide empirical evidence for space and time (see ``Physics'' section).

\subsubsection{Local Rule}
To apply Lenia's local rule to every site $\mathbf{x}$ at time $t$, concolve the grid with a \textit{kernel} $\mathbf{K}: \mathcal{N} \rightarrow \mathcal{S}$ to yield the \textit{potential distribution} $\mathbf{U}^t$:
\begin{equation}
\mathbf{U}^t(\mathbf{x}) = \mathbf{K} * \mathbf{A}^t(\mathbf{x}) =
\begin{cases}
\displaystyle\sum_{n\in\mathcal{N}} \mathbf{K}(\mathbf{n}) \mathbf{A}^t(\mathbf{x}+\mathbf{n}) \; \Delta x^2 & \text{ in DL} \\
\displaystyle\int_{n \in \mathcal{N}} \mathbf{K}(\mathbf{n}) \mathbf{A}^t(\mathbf{x}+\mathbf{n}) \; \mathrm{d}x^2 & \text{ in CL}
\end{cases}
\end{equation}
Feed the potential into a \textit{growth mapping} $G: [0, 1] \rightarrow [-1, 1]$ to yield the \textit{growth distribution} $\mathbf{G}^t$:
\begin{equation}
\mathbf{G}^t(\mathbf{x}) = G(\mathbf{U}^t(\mathbf{x}))
\end{equation}
Update every site by adding a small fraction $\Delta t$ ($\mathrm{d}t$ in CL) of the growth and clipping back to the unit range $[0, 1]$; the time is now $t+\Delta t$:
\begin{equation}
\mathbf{A}^{t+\Delta t}(\mathbf{x}) = \left[\mathbf{A}^t(\mathbf{x}) + \Delta t \; \mathbf{G}^t(\mathbf{x}) \right]_0^1
\end{equation}
where $[n]_a^b = \min(\max(n, a), b)$ is the clip function.

\subsubsection{Kernel}
The kernel $\mathbf{K}$ is constructed by \textit{kernel core} $K_C: [0, 1] \rightarrow [0, 1]$ which determines its detailed ``texture'', and \textit{kernel shell} $K_S: [0, 1] \rightarrow [0, 1]$ which determines its overall ``skeleton'' (Figure \ref{fig-core}(a-e)).

The kernel core $K_C$ is any unimodal function satisfying $K_C(0) = K_C(1) = 0$ and usually $K_C(\frac{1}{2}) = 1$.  By taking polar distance as argument, it creates a uniform ring around the site:
\begin{equation}
K_C(r) =
\begin{cases}
\exp\left(\alpha - \displaystyle\frac{\alpha}{4r(1-r)}\right) & \text{ exponential, } \alpha = 4 \\
\left(4r(1-r)\right)^\alpha & \text{ polynomial, } \alpha = 4 \\
\mathbf{1}_{[\frac{1}{4}, \frac{3}{4}]}(r) & \text{ rectangular} \\
\ldots & \text{ or others}
\end{cases}
\end{equation}
where $\mathbf{1}_{A}(r)$ = 1 if $r \in A$ else 0 is the indicator function.

The kernel shell $K_S$ takes a parameter vector $\beta = (\beta_1, \beta_2, \ldots, \beta_B) \in [0, 1]^B$ (\textit{kernel peaks}) of size $B$ (\textit{the rank}) and copies the kernel core into equidistant concentric rings with peak heights $\beta_i$:
\begin{equation}
K_S(r; \beta) = \beta_{\lfloor Br \rfloor} K_C(Br \text{ mod } 1)
\end{equation}
Finally, the kernel is normalized to makes sure $\mathbf{K} * \mathbf{A} \in [0, 1]$:
\begin{equation}
\mathbf{K}(\mathbf{n}) = \frac{K_S(\|\mathbf{n}\|_2)}{|K_S|}
\end{equation}
where $|K_S| = \sum_{\mathcal{N}} K_S \; \Delta x^2$ in DL, or $\int_{\mathcal{N}} K_S \; \mathrm{d}x^2$ in CL.

Notes on parameter $\beta$:

\begin{itemize}[noitemsep]
\item To compare $\beta$ of different ranks, a vector $\beta$ is equivalent to one with $n$ trailing zeros while space resolution $R$ is scaled by $(B+n)/B$ at the same time, e.g. $\beta = (1) \equiv (1, 0, 0)$ with $R$ scaled by 3.
\item To compare $\beta$ of the same rank, a vector $\beta$ where $\forall i \; \beta_i \neq 1$ is equivalent to a scaled one $\beta / \max(\beta_i)$ such that $\exists i \; \beta_i = 1$ while the kernel remains unchanged due to normalization, e.g. $\beta = (\frac{1}{3}, 0, \frac{2}{3}) \equiv (\frac{1}{2}, 0, 1)$.
\item Consequently, all possible $\beta$ as a $B$-dimensional hypercube can be projected onto its $(B-1)$-dimensional hypersurfaces.  (see Figure \ref{fig-cube})
\end{itemize}

\subsubsection{Growth Mapping}
The growth mapping $G: [0, 1] \rightarrow [-1, 1]$ is any unimodal, nonmonotonic function with parameters $\mu, \sigma \in \mathbb{R}$ (\textit{growth center} and \textit{growth width}) satisfying $G(\mu) = 1$ (cf. $\zeta(\cdot)$ in \cite{MacLennan1990}) (Figure \ref{fig-core}(f-g)):
\begin{equation}
G(u; \mu, \sigma) =
\begin{cases}
2 \; \exp\left(- \displaystyle\frac{(u - \mu)^2}{2\sigma^2}\right) - 1 & \text{ exponential} \\
2 \; \mathbf{1}_{[\mu \pm 3 \sigma]}(u) \left(1 - \displaystyle\frac{(u - \mu)^2}{9\sigma^2}\right)^\alpha - 1 & \text{ polynomial, } \alpha = 4 \\
2 \; \mathbf{1}_{[\mu \pm \sigma]}(u) - 1 & \text{ rectangular} \\
\ldots & \text{ or others}
\end{cases}
\end{equation}

\subsubsection{GoL inside Lenia}

GoL can be considered a special case of discrete Lenia with $R = T = P = 1$, using a variant of the rectangular kernel core:
\begin{equation}
K_C(r) = \mathbf{1}_{[\frac{1}{4}, \frac{3}{4}]}(r) + \frac{1}{2} \; \mathbf{1}_{[0, \frac{1}{4})}(r)
\end{equation}
and the rectangular growth mapping with $\mu=0.35, \sigma=0.07$.

\subsubsection{Summary}
In summary, discrete and continuous Lenia are defined as:
\begin{align}
\mathcal{A}_\text{DL} &= \left(\Delta x \; \mathbb{Z}^2, \Delta t \; \mathbb{Z}, \Delta p \; \{0 \ldots P\}, \mathbb{B}_1[0], \right.\nonumber \\
&\qquad \left. \mathbf{A}^{t+\Delta t} \mapsto \left[\mathbf{A}^t + \Delta t \; G_{\mu,\sigma}(\mathbf{K}_{\beta} * \mathbf{A}^t)\right]_0^1 \right) \\
\mathcal{A}_\text{CL} &= \left(\mathbb{R}^2, \mathbb{R}, [0, 1], \mathbb{B}_1[0], \right.\nonumber \\
&\qquad \left. \mathbf{A}^{t+\mathrm{d}t} \mapsto \left[\mathbf{A}^t + \mathrm{d} t \; G_{\mu,\sigma}(\mathbf{K}_{\beta} * \mathbf{A}^t)\right]_0^1 \right)
\end{align}

The associated dimensions are: space-time-state resolutions $R, T, P$, steps $\Delta x, \Delta t, \Delta p$, infinitesimals $\mathrm{d}x, \mathrm{d}t, \mathrm{d}p$.  The associated parameters are: growth center $\mu$, growth width $\sigma$, kernel peaks $\beta$ of rank $B$.  The mutable core functions are: kernel core $K_C$, growth mapping $G$.

\subsection{Computer Implementation}

Discrete Lenia (DL) can be implemented with the pseudocode below, assuming an array programming language is used (e.g. Python with NumPy, MATLAB, Wolfram).

\begin{figure}[!tb]
\centering
\includegraphics[width=\textwidth,trim=4 4 4 4,clip]{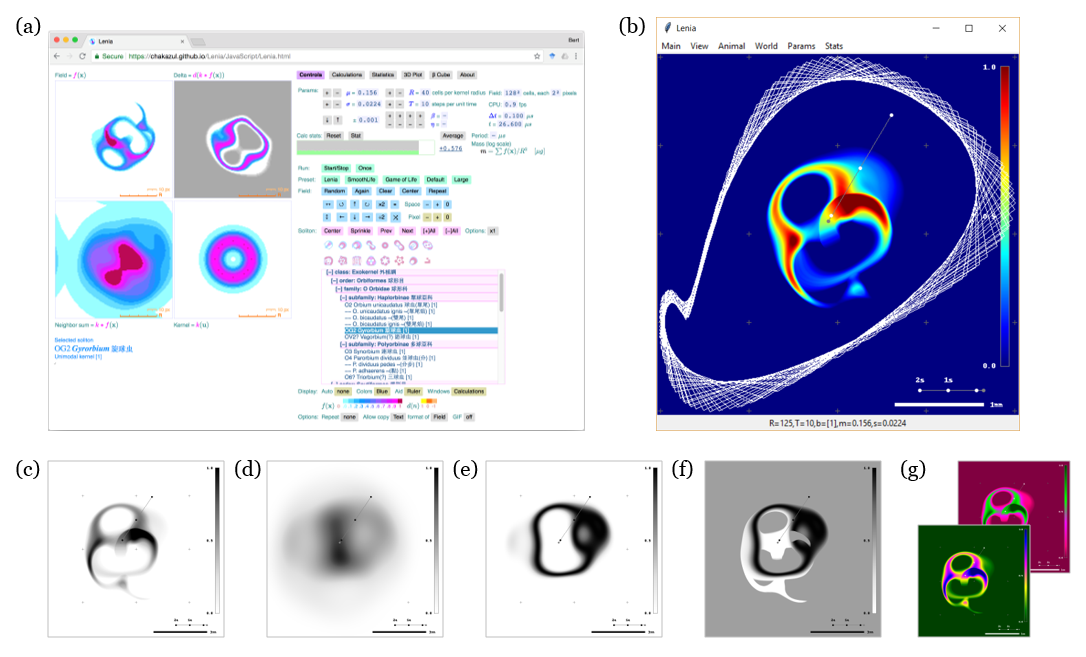}
\caption{Computer implementations of Lenia with interactive user interfaces.  \textbf{(a-b)} Web version run in Chrome browser (a) and Python version with GPU support (b).  \textbf{(c-f)} Different views during simulation, including the configuration $\mathbf{A}^t$ (c), the potential $\mathbf{U}^t$ (d), the growth $\mathbf{G}^t$ (e), and the actual change $\Delta \mathbf{A}/\Delta t = (\mathbf{A}^{t+\Delta t} - \mathbf{A}^t)/\Delta t$ (f).  \textbf{(g)} Other color schemes.}
\label{fig-impl}
\end{figure}

Interactive programs have been written in JavaScript / HTML5, Python, and MATLAB to provide user interface for new species discovery (Figure \ref{fig-impl}(a-b)).  Non-interactive program has been written in C\#.NET for automatic traverse through the parameter space using a flood fill algorithm (breath-first or depth-first search), providing species distribution, statistical data and occasionally new species.

State precision $\Delta p$ can be implicitly implemented as the precision of floating-point numbers.  For values in the unit interval $[0, 1]$, the precision ranges from $2^{-126}$ to $2^{-23}$ (about $1.2 \times 10^{-38}$ to $1.2 \times 10^{-7}$) using 32-bit single-precision, or from $2^{-1022}$ to $2^{-52}$ (about $2.2 \times 10^{-308}$ to $2.2 \times 10^{-16}$) using 64-bit double-precision \cite{Kahan1997}.  That means $P > 10^{15}$ using double precision.

Discrete convolution can be calculated as the sum of element-wise products:
\begin{equation}
\mathbf{K} * \mathbf{A}^t(\mathbf{x}) = \sum_{\mathbf{n}\in\mathcal{N}} \mathbf{K}(\mathbf{n}) \mathbf{A}^t(\mathbf{x}+\mathbf{n})
\end{equation}
or alternatively, using discrete Fourier transform (DFT) according to the convolution theorem:
\begin{equation}
\mathbf{K} * \mathbf{A}^t = \mathcal{F}^{-1} \left\{ \mathcal{F} \{ \mathbf{K} \} \cdot \mathcal{F} \{ \mathbf{A}^t \} \right\}
\end{equation}

Efficient calculation can be achieved using fast Fourier transform (FFT) \cite{CooleyTukey1965}, pre-calculation of the kernel's FFT $\mathcal{F}\{\mathbf{K}\}$, and parallel computing like GPU acceleration.  The DFT/FFT approach automatically produces a periodic boundary condition.

\subsubsection{Pseudocode}
Symbol \texttt{@} indicates two-dimensional matrix of floating-point numbers.

\begin{footnotesize}
\begin{verbatim}
function pre_calculate_kernel(beta, dx)
  @radius = get_polar_radius_matrix(SIZE_X, SIZE_Y) * dx
  @Br = size(beta) * @radius
  @kernel_shell = beta[floor(@Br)] * kernel_core(@Br % 1)
  @kernel = @kernel_shell / sum(@kernel_shell)
  @kernel_FFT = FFT_2D(@kernel)
  return @kernel, @kernel_FFT
end

function run_automaton(@world, @kernel, @kernel_FFT, mu, sigma, dt)
  if size(@world) is small
    @potential = elementwise_convolution(@kernel, @world)
  else
    @world_FFT = FFT_2D(@world)
    @potential_FFT = elementwise_multiply(@kernel_FFT, @world_FFT)
    @potential = FFT_shift(real_part(inverse_FFT_2D(@potential_FFT)))
  end
  @growth = growth_mapping(@potential, mu, sigma)
  @new_world = clip(@world + dt * @growth, 0, 1)
  return @new_world, @growth, @potential
end

function simulation()
  R, T, mu, sigma, beta = get_parameters()
  dx = 1/R;  dt = 1/T;  time = 0
  @kernel, @kernel_FFT = pre_calculate_kernel(beta, dx)
  @world = get_initial_configuration(SIZE_X, SIZE_Y)
  repeat
    @world, @growth, @potential = run_automaton(@world, 
        @kernel, @kernel_FFT, mu, sigma, dt)
    time = time + dt
    display(@world, @potential, @growth)
  end
end
\end{verbatim}
\end{footnotesize}

\subsubsection{User Interface}
For implementations requiring an interactive user interface, one or more of the following components are recommended:

\begin{itemize}[noitemsep]
\item Controls for starting and stopping CA simulation
\item Panels for displaying different stages of CA calculation
\item Controls for changing parameters and space-time-state resolutions
\item Controls for randomizing, transforming and editing the configuration
\item Controls for saving, loading, and copy-and-pasting configurations
\item Clickable list for loading predefined patterns
\item Utilities for capturing the display output (e.g. image, GIF, movie)
\item Controls for customizing the layout (e.g. grid size, color map)
\item Controls for auto-centering, auto-rotating and temporal sampling
\item Panels or overlays for displaying real-time statistical analysis
\end{itemize}

\subsubsection{Pattern Storage}
A pattern can be stored for publication and sharing using a data exchange format (e.g. JSON, XML) that includes the \textit{run-length encoding} (RLE) of the two-dimensional array $\mathbf{A}^t$ and its associated settings $(R, T, P, \mu, \sigma, \beta, K_C, G)$, or alternatively, using a plaintext format (e.g. CSV) for further analysis or manipulation in numeric software.

A long list of interesting patterns can be saved as JSON/XML for program retrieval.  To save storage space, patterns can be stored with space resolution $R$ as small as possible (usually $10 \leq R \leq 20$) thanks to Lenia's scale invariance (see ``Physics'' section).

\subsubsection{Environment}
Most of computer simulations, experiments, statistical analysis, image and video capturing for this paper were done using the following environments and settings:

\begin{itemize}[noitemsep]
\item Hardware: Apple MacBook Pro (OS X Yosemite), Lenovo ThinkPad X280 (Microsoft Windows 10 Pro)
\item Software: Python 3.7.0, MathWorks MATLAB Home R2017b, Google Chrome browser, Microsoft Excel 2016
\item State precision: double precision
\item Kernel core and growth mapping: exponential
\end{itemize}

\subsection{Evolving New Species}

A self-organizing, autonomous pattern in Lenia is called a \textit{lifeform}, and a kind of similar lifeforms is called a \textit{species}.  Up to the moment, more than 400 species have been discovered.  \textit{Interactive evolutionary computation} (IEC) \cite{Takagi2001} is the major force behind the generation, mutation and selection of new species.  In evolutionary computation (EC), the fitness function is usually well known and can be readily calculated.  However, in the case of Lenia, due to the non-trivial task of pattern recognition, as well as aesthetic factors, evolution of new species often requires human interaction.

Interactive computer programs provide user interface and utilities for human users to carry out mutation and selection operators manually.  Mutation operators include parameter tweaking and configuration manipulation.  Selection operators include observation via different views for fitness estimation (Figure \ref{fig-core}(c-f)) and storage of promising patterns.  Selection criteria include survival, long-term stability, aesthetic appeal, and novelty.

Listed below are a few evolutionary strategies learnt from experimenting and practicing.

\begin{figure}[!tb]
\centering
\includegraphics[width=\textwidth,trim=4 4 4 4,clip]{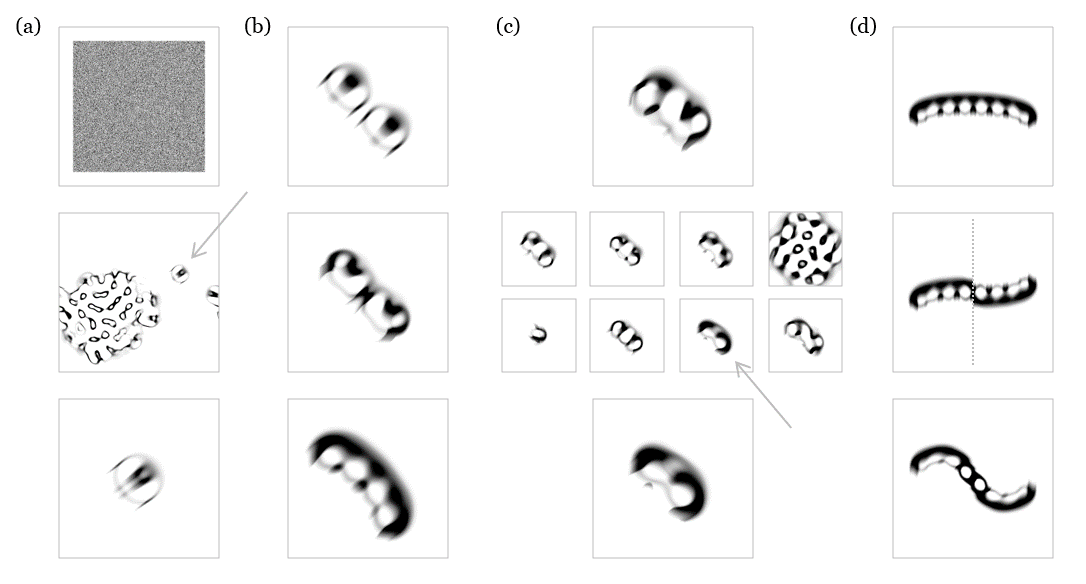}
\caption{Strategies of evolving new Lenia lifeforms using Interactive Evolutionary Computation (IEC).  \textbf{(a)} Random generation: random initial configuration is generated (top) and simulation is run (middle), where new lifeforms were spotted (arrow) and isolated (bottom).  \textbf{(b)} Parameter tweaking: with an existing lifeform (top), parameters are adjusted so that new morphologies or behaviors are observed (middle, bottom).  \textbf{(c)} Automatic exploration: a starting lifeform (top) is put into an automatic program to explore wide ranges of parameters (middle), where new lifeforms were occasionally discovered (arrow) and isolated (bottom).  \textbf{(d)} Manual mutation: an existing lifeform (top) is modified, here single-side flipped (middle), and parameter tweaked to stabilize into a new species (bottom).}
\label{fig-evo}
\end{figure}

\subsubsection{Random Generation}
Initial configurations with random patches of non-zero sites were generated and put into simulation using interactive program.  This is repeated using different random distributions and different parameters.  Given enough time, naturally occurring lifeforms would emerge from the primordial soup, for example \textit{Orbium}, \textit{Scutium}, \textit{Paraptera}, and radial symmetric patterns.  (Figure \ref{fig-evo}(a))

\subsubsection{Parameter Tweaking}
Using an existing lifeform, parameters were changed progressively or abruptly, forcing the lifeform to die out (explode or evaporate) or survive by changing slightly or morphing into another species.  Any undiscovered species with novel structure or behavior were recorded.  (Figure \ref{fig-evo}(b))

Transient patterns captured during random generation could also be stabilized into new species in this way.

Long-chain lifeforms (e.g. \textsf{Pterifera}) could first be elongated by temporary increasing the growth rate (decrease $\mu$ or increase $\sigma$), then stabilized into new species by reversing growth.  Shortening could be done in the opposite manner.

\subsubsection{Automatic Exploration}
Starting from an existing lifeform, automatic program was used to traverse the parameter space (i.e. continuous parameter tweaking).  All survived patterns were recorded, among them new species were occasionally found.  Currently, automated exploration is ineffective without the aid of artificial intelligence (e.g. pattern recognition), and has only been used for simple conditions (rank 1, mutation by parameter tweaking, selection by survival).  (Figure \ref{fig-evo}(c))

\subsubsection{Manual Mutation}
Patterns were edited or manipulated (e.g. enlarging, shrinking, mirroring, single-side flipping, recombining) using our interactive program or other numeric software, and then parameter tweaked in attempt to stabilize into new species.  (Figure \ref{fig-evo}(d))

\subsection{Analysis of Lifeforms}

\subsubsection{Qualitative Analysis}
By using computer simulation and visualization and taking advantage of human's innate ability of spatial and temporal pattern recognition, the physical appearances and movements of known species were being observed, documented and categorized, as reported in the ``Morphology'' and ``Behavior'' sections.  Using automatic traverse program, the distributions of selected species in the parameter space were charted, as reported in the ``Ecology'' section.  A set of criteria, based on the observed similarities and differences among the known species, were devised to categorize them into a hierarchical taxonomy, as reported in the ``Taxonomy'' section.

\subsubsection{Quantitative Analysis}
Statistical methods were used to analyze lifeforms to compensate the limitations in human observation regarding subtle variations and long-term trends.  A number of \textit{statistical measures} were calculated over the configuration (i.e. mass distribution) $\mathbf{A}$ and the positive-growth distribution $\mathbf{G}|_{\mathbf{G}>0}$:

\begin{itemize}[noitemsep]
\item \textit{Mass} is the sum of states, $m = \int \mathbf{A}(\mathbf{x}) \mathrm{d}\mathbf{x} \;  [\text{mg}]$
\item \textit{Volume} is the number of positive states, $V_m = \int_{\mathbf{A}>0} \mathrm{d}\mathbf{x} \;  [\text{mm}^2]$
\item \textit{Density} is the overall density of states, $\rho_m = m / V_m\;  [\text{mg mm}^{-2}]$
\item \textit{Growth} is the sum of positive growth, $g = \int_{\mathbf{G}>0} \mathbf{G}(\mathbf{x}) \mathrm{d}\mathbf{x} \;  [\text{mg s}^{-1}]$
\item \textit{Centroid} is the center of states, $\bar{\mathbf{x}}_m = \int \mathbf{x} \mathbf{A}(\mathbf{x}) \mathrm{d}\mathbf{x} / m$
\item \textit{Growth center} is the center of positive growth, $\bar{\mathbf{x}}_g = \int_{\mathbf{G}>0} \mathbf{x} \mathbf{G}(\mathbf{x}) \mathrm{d}\mathbf{x} / g$
\item \textit{Growth-centroid distance} is the distance between the two centers, \\
$d_{gm} = |\bar{\mathbf{x}}_g - \bar{\mathbf{x}}_m| \;  [\text{mm}]$
\item \textit{Linear speed} is the linear moving rate of the centroid, \\
$s_m = |\mathrm{d} \bar{\mathbf{x}}_m / \mathrm{d}t| \;  [\text{mm s}^{-1}]$
\item \textit{Angular speed} is the angular moving rate of the centroid, \\
$\omega_m = \mathrm{d} / \mathrm{d}t \arg(\mathrm{d} \bar{\mathbf{x}}_m / \mathrm{d}t) \;  [\text{rad s}^{-1}]$
\item \textit{Mass asymmetry} is the mass difference across the directional vector, \\
$m_\Delta = \int_{c>0} \mathbf{A}(\mathbf{x}) \mathrm{d}\mathbf{x} - \int_{c<0} \mathbf{A}(\mathbf{x}) \mathrm{d}\mathbf{x} \;  [\text{mg}]$ where $c = \mathrm{d}\bar{\mathbf{x}}_m \times (\mathbf{x} - \bar{\mathbf{x}}_m)$
\item \textit{Angular mass} is the second moment of mass from the centroid, \\
$I_m = \int \mathbf{A}(\mathbf{x}) (\mathbf{x} - \bar{\mathbf{x}}_m)^2 \mathrm{d}\mathbf{x} \;  [\text{mg mm}^2]$
\item \textit{Gyradius} is the root-mean-square of site distances from the centroid, \\
$r_m = \sqrt{I_m / m} \;  [\text{mm}]$
\item Others e.g. Hu's and Flusser's moment invariants $\phi_i$ \cite{Hu1962, Flusser2006}
\end{itemize}

Note: Brackets indicate the units of measure borrowed from SI units in microscopic scale, e.g. ``mm'' for length, ``rad'' for angle, ``s'' for time, ``mg'' for states (cf. ``lu'' and ``tu'' in \cite{Munafo2014}).

Based on the multivariate time-series of statistical measures, the following ``meta-measures'' could be calculated:

\begin{itemize}[noitemsep]
\item Summary statistics (mean, median, standard deviation, minimum, maximum, quartiles)
\item Quasi-period (estimated using e.g. autocorrelation, periodogram)
\item Degree of chaos (e.g. Lyapunov exponent, attractor dimension)
\item Probability of survival
\end{itemize}

The following charts were plotted using various parameters, measures and meta-measures:

\begin{itemize}[noitemsep]
\item Time series chart (measure vs. time)
\item Phase space trajectory (measure vs. measure) (e.g. Figure \ref{fig-cross} insets)
\item Allometric chart (meta vs. meta) (e.g. Figure \ref{fig-mea}, \ref{fig-u4mea})
\item Cross-sectional chart (meta vs. parameter) (e.g. Figure \ref{fig-cross})
\item $\mu$-$\sigma$ map (parameter $\mu$ vs. $\sigma$; information as color) (e.g. Figure \ref{fig-map}, \ref{fig-u4map})
\item $\beta$-cube (parameter $\beta$ components as axes; information as color) (e.g. Figure \ref{fig-cube})
\end{itemize}

Over 1.2 billion measures were collected using automatic traverse program and analyzed using numeric software like Microsoft Excel.  Results are presented in the ``Physiology'' section.

\subsubsection{Spatiotemporal Analysis}
Constant motions like translation, rotation and oscillation render visual analysis difficult.  It is desirable to separate the spatial and temporal aspects of a moving pattern so as to directly assess the static form and estimate the motion frequencies (or quasi-periods).

Linear motion can be removed by \textit{auto-centering}, to display the pattern centered at its centroid $\bar{\mathbf{x}}_m$.

Using \textit{temporal sampling}, the simulation is displayed one frame per $N$ time-steps.  When any rotation is perceived as near stationary due to the stroboscopic effect, the rotation frequency is approximately the sampling frequency $f_r \approx f_s = 1/(N\Delta t)$.  Calculate the sampled angular speed $\omega_s = \theta f_r = 2 \pi f_r/n$ where $n$ is the number of radial symmetric axes.  Angular motion can be removed by \textit{auto-rotation}, to display the pattern rotated by $-\omega_s t$.

With the non-translating, non-rotating pattern, any global or local oscillation frequency can be determined as $f_o \approx f_s$ again using temporal sampling.

\section{RESULTS}

Results of the study of Lenia will be outlined in various sections: Physics, Taxonomy, Ecology, Morphology, Behavior, Physiology, and Case Study.

\subsection{Physics}

We present general results regarding the effects of basic CA settings, akin to physics where one studies how the space-time fabric and fundamental laws influence matter and energy.
 
\begin{figure}[!tb]
\centering
\includegraphics[width=\textwidth,trim=4 4 4 4,clip]{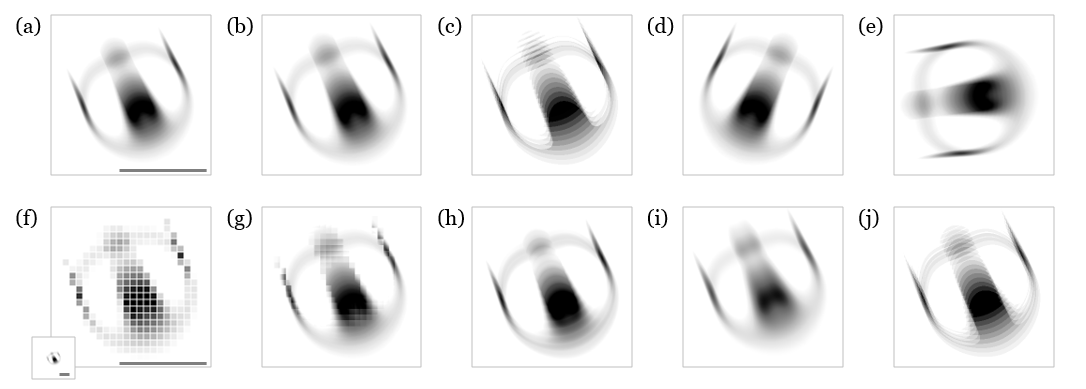}
\caption{
Plasticity of \textit{Orbium} ($\mu=0.15, \sigma=0.016$) under various environment settings and transformations.  (Scale bar is unit length = kernel radius, same in all panels).  \textbf{(a)} Original settings: $R=185, T=10, P>10^{15}$ (double precision), exponential core functions.  \textbf{(b-c)} Core functions changed to polynomial with no visible effect (b), to rectangular produces rougher pattern (c).  \textbf{(d-e)} Pattern flipped horizontally (d) or rotated 77\degree anti-clockwise (e) with no visible effect.  \textbf{(f-g)} Pattern downsampled with space compressed to $R=15$ (f: zoomed in, inset: actual size), under recovery after upsampled using nearest-neighbor and space resolution restored to $R=185$ (g), eventually recovers to (a).  \textbf{(h-i)} Time compressed to $T=5$ produces rougher pattern (h); time dilated to $T=320$ produces smoother, lower density pattern (i).  \textbf{(j)} Fewer states $P=10$ produces rougher pattern.
}
\label{fig-phy}
\end{figure}

\begin{figure}[!tb]
\centering
\includegraphics[width=\textwidth,trim=4 4 4 4,clip]{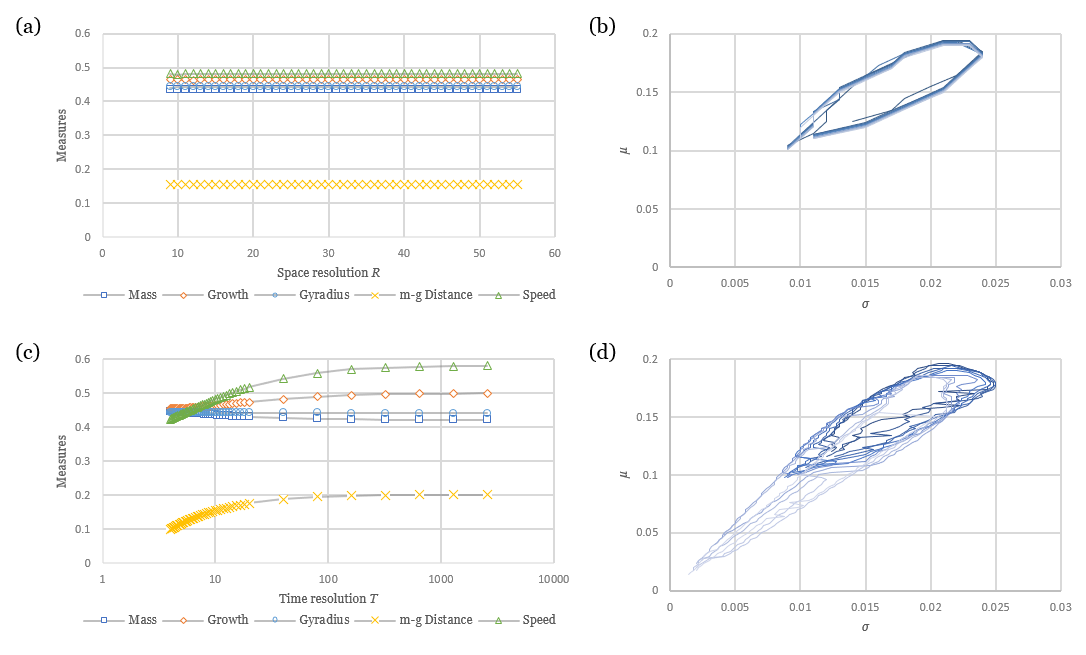}
\caption{Effects of space-time resolutions as experimented with \textit{Orbium} ($\mu=0.15, \sigma=0.016$).  Each data point in (a) and (c) is averaged across 300 time-steps.  \textbf{(a-b)} Spatial invariance: for a range of space resolution $R \in \{9 \ldots 55\}$ and fixed time resolution T=10, all statistical measures (mass $m$, growth $g$, gyradius $r_m$, growth-centroid distance $d_{gm}$, linear speed $s_m$) remain constant (a) and the parameter range (``niche'') remain static (total 557 loci) (b).  \textbf{(c-d)} Temporal asymptosy: for a range of time resolution $T \in \{4 \ldots 2560\}$ and fixed space resolution R=13, structure-related measures ($m$, $r_m$) go down and dynamics-related measures ($g$, $d_{gm}$, $s_m$) go up, reaching each continuum limit asymptically (c); the parameter range expands as time dilates (dark to light enclosures, total 14,182 loci) (d).}
\label{fig-res}
\end{figure}

\subsubsection{Spatial Invariance}
For sufficiently fine space resolution ($R > 12$), patterns in Lenia are minimally affected by spatial similarity transformations including shift, rotation, reflection and scaling (Figure \ref{fig-phy}(d-g)).  Shift invariance is shared by all homogenous CAs; reflection invariance is enabled by symmetries in neighborhood and local rule; scale invariance is enabled by large neighborhoods (as in LtL \cite{Evans2001}); rotation invariance is enabled by circular neighborhoods and totalistic or polar local rules (as in SmoothLife \cite{Rafler2011} and Lenia).  Our empirical data of near constant metrics of \textit{Orbium} over various space resolutions $R$ further supports scale invariance in Lenia (Figure \ref{fig-res}(a-b)).

\subsubsection{Temporal Asymptosy}
The local rule $\phi$ of discrete Lenia (DL) can be considered the Euler method $\mathbf{A}_{n+1} = \mathbf{A}_n + h f(\mathbf{A}_n)$ for solving the local rule $\phi$ of continuous Lenia (CL) rewritten as an ordinary differential equation (ODE):
\begin{align}
\mathbf{A}^{t+dt} &= \mathbf{A}^t + \mathrm{d}t \left[ G(\mathbf{K} * \mathbf{A}^t) \right]_{-\mathbf{A}^t/\mathrm{d}t}^{(1-\mathbf{A}^t)/\mathrm{d}t} \\
\frac{\mathrm{d}}{\mathrm{d}t} \mathbf{A}^t &= \left[ G(\mathbf{K} * \mathbf{A}^t) \right]_{-\mathbf{A}^t/\mathrm{d}t}^{(1-\mathbf{A}^t)/\mathrm{d}t}
\end{align}

The Euler method should better approximate the ODE as step size $h$ diminishes, similarly DL should approach its continuum limit CL as $\Delta t$ decreases.  This is supported by empirical data of asymptotic metrics of \textit{Orbium} over increasing time resolutions $T$ (Figure \ref{fig-res}(c-d)) towards an imaginable ``true \textit{Orbium}'' (Figure \ref{fig-phy}(i)).

\subsubsection{Core Functions}
Choices of kernel core $K_C$ and growth mapping $G$ (the core functions or ``fundamental laws'') usually alter the ``textures'' of a pattern but not its overall structure and dynamics (Figure \ref{fig-phy}(b-c)).  Smoother core functions (e.g. exponential) produce smoother patterns, rougher ones (e.g. rectangular) produce rougher patterns.  This plasticity suggests that similar lifeforms should exist in SmoothLife which resembles Lenia with rectangular core functions, as supported by similar creatures found in both CAs (Figure \ref{fig-alife}(f-g)).

\subsection{Taxonomy}

We present the classification of Lenia lifeforms into a hierarchical taxonomy, a process comparable to the biological classification of Terrestrial life \cite{Linnaeus1758}.

\subsubsection{Phylogeny of the Glider}
The most famous moving pattern in GoL is the diagonally-moving ``glider'' (Figure \ref{fig-alife}(a)).  It was not until LtL \cite{Evans2003} that scalable digital creatures were discovered including the glider analogue ``bugs with stomach'', and SmoothLife \cite{Rafler2011} was the first to produce an omni-directional bug called the ``smooth glider'', which was rediscovered in Lenia as \textit{Scutium} plus variants (Figure \ref{fig-alife}(e-g) left).  We propose the phylogeny of the glider:
\begin{quote}
Glider $\rightarrow$ Bug with stomach $\rightarrow$ Smooth glider $\rightarrow$ \textit{Scutium}
\end{quote}
Phylogenies of other creatures are possible, like the ``wobbly glider'' and \textit{Pyroscutium} (Figure \ref{fig-alife}(f-g) right).

\subsubsection{Classification}
Principally there are infinitely many types of lifeforms in Lenia, but a range of visually and statistically similar lifeforms were grouped into a \textit{species}, defined such that one \textit{instance} can be morphed smoothly into another by continuously adjusting parameters or other settings.
Species were further grouped into higher \textit{taxonomic ranks} --- genera, families, orders, classes --- with decreasing similarity and increasing generality, finally subsumed into phylum \textsf{Lenia}, kingdom \textsf{Automata}, domain \textsf{Simulata}, and the root \textsf{Artificialia}.  Potentially other kinds of artificial life can be incorporated into this \textsf{Artificialia} tree.

\begin{figure}[!t]
\centering
\includegraphics[width=\textwidth,trim=4 4 4 4,clip]{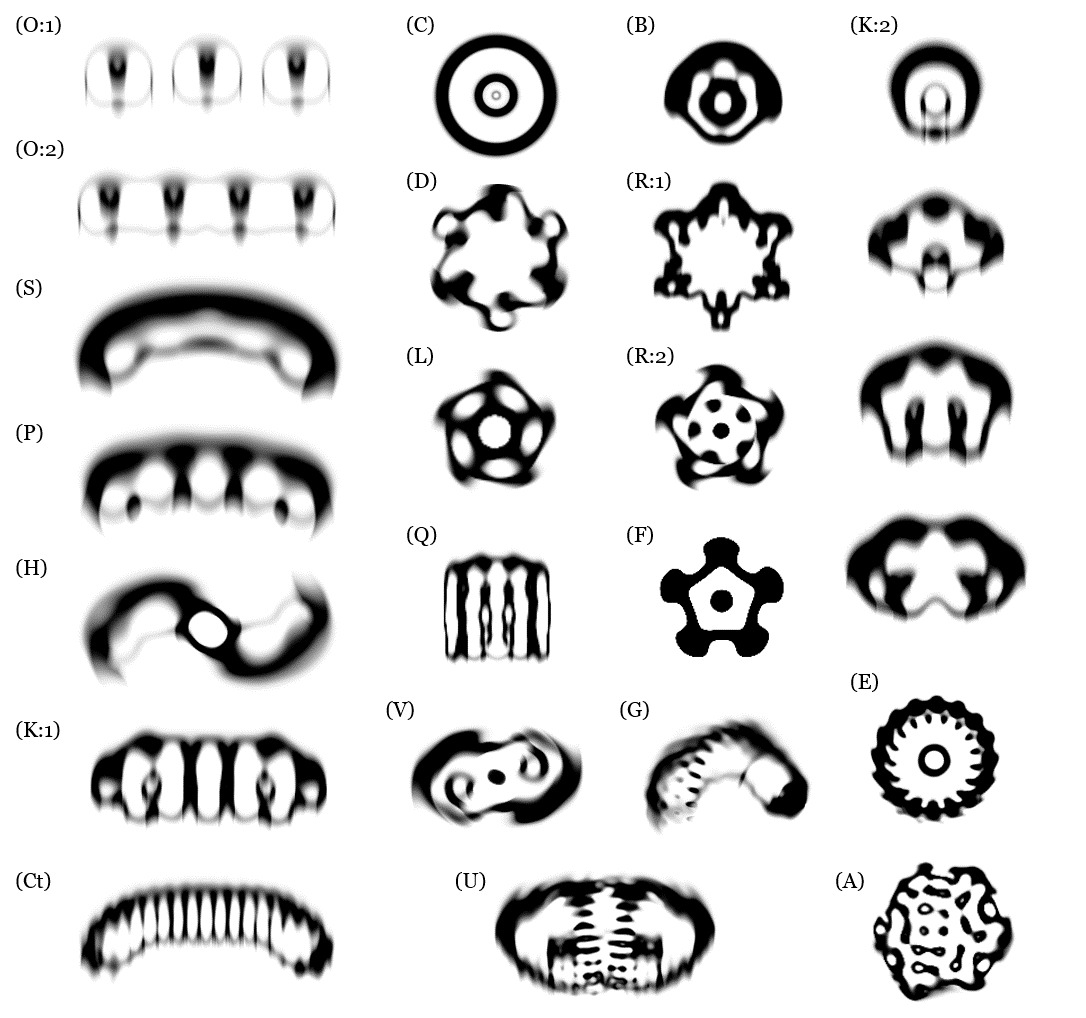}
\caption{Biodiversity in Lenia as exemplified by the 18 Lenia families (not to scale).  \textbf{(Column 1)} \textsf{(O) Orbidae, (S) Scutidae, (P) Pterifera, (H) Helicidae, (K) Kronidae, (Ct) Ctenidae}; \textbf{(Column 2)} \textsf{(C) Circidae, (D) Dentidae, (L) Lapillidae, (Q) Quadridae, (V) Volvidae}; \textbf{(Column 3)} \textsf{(B) Bullidae, (R) Radiidae, (F) Folidae, (G) Geminidae, (U) Uridae}; \textbf{(Column 4)} \textsf{(K) Kronidae, (E) Echinidae, (A) Amoebidae}.}
\label{fig-tax}
\end{figure}

Below are the current definitions of the taxonomic ranks.

\begin{itemize}[noitemsep]
\item A \textit{species} is a group of lifeforms with the same morphology and behavior in global and local scales, form a cluster (niche) in the parameter space, and follow the same statistical trends in the phase space (Figure \ref{fig-map}, \ref{fig-mea}).  Continuous morphing among members is possible.
\item A \textit{genus} is a group of species with the same global morphology and behavior but differ locally, occupy adjacent niches, and have discontinuity in statistical trends.  Abrupt but reversible transformation among member species is possible.
\item A \textit{subfamily} is a series of genera with increasing number of ``units'' or ``vacuoles'', occupy parallel niches of similar shapes.
\item A \textit{family} is a collection of subfamilies with the same architecture or body plan, composed of the same set of components arranged in similar ways.
\item An \textit{order} is a rough grouping of families with similar architectures and statistical qualities, e.g. speed.
\item A \textit{class} is a high-level grouping of how lifeforms influenced by the arrangement of kernel.
\end{itemize}

\subsubsection{Tree of Artificial Life}

The notion of ``life'', here interpreted as self-organizing autonomous entities in a broader sense, may include biological life, artificial life, and other possibilities like extraterrestrial life.  Based on lifeforms from Lenia and other systems, we propose the \textit{tree of artificial life}: \\

{\small
\noindent \textsf{Artificialia} \\
\indent Domain \textsf{Synthetica} \tabto{4.5cm} ``Wet'' biochemical synthetic life \\
\indent Domain \textsf{Mechanica} \tabto{4.5cm} ``Hard'' mechanical or robotic life, e.g. \cite{Jansen2008} \\
\indent Domain \textsf{Simulata} \tabto{4.5cm} ``Soft'' computer simulated life \\
\indent \indent Kingdom \textsf{Sims} \tabto{4.5cm} Evolved virtual creatures, e.g. \cite{Sims1994, Cheney2013, Kriegman2018} \\
\indent \indent Kingdom \textsf{Greges} \tabto{4.5cm} Particle swarm solitons, e.g. \cite{Schmickl2016, Sayama2018b, Sayama2011, Sayama2018} \\
\indent \indent Kingdom \textsf{Turing} \tabto{4.5cm} Reaction-diffusion solitons, e.g. \cite{Munafo2014, Munafo2009, Hutton2012b} \\
\indent \indent Kingdom \textsf{Automata} \tabto{4.5cm} Cellular automata solitons \\
\indent \indent \indent Phylum \textsf{Discreta} \tabto{4.5cm} Non-scalable, e.g. \cite{Adamatzky2010, LifeWiki, Cook2004} \\
\indent \indent \indent Phylum \textsf{Lenia} \tabto{4.5cm} Scalable, e.g. \cite{Evans2001, Rafler2011} \\
}

The current taxonomy of Lenia (Figure \ref{fig-tax}): \\

{\small
\noindent Phylum \textsf{Lenia} \\
\indent Class \textsf{Exokernel} \tabto{4.6cm} having strong outer kernel rings \\
\indent \indent Order \textsf{Orbiformes}  \\
\indent \indent \indent Family \textsf{Orbidae (O)} \tabto{4.6cm} ``disk bugs'', disks with central stalk \\
\indent \indent Order \textsf{Scutiformes} \\
\indent \indent \indent Family \textsf{Scutidae (S)} \tabto{4.6cm} ``shield bugs'', disks with thick front \\
\indent \indent \indent Family \textsf{Pterifera (P)} \tabto{4.6cm} ``winged bugs'', one/two wings with sacs \\
\indent \indent \indent Family \textsf{Helicidae (H)} \tabto{4.6cm} ``helix bugs'', rotating versions of \textsf{P} \\
\indent \indent \indent Family \textsf{Circidae (C)} \tabto{4.6cm} ``circle bugs'', one or more concentric rings \\
\indent  Class \textsf{Mesokernel} \tabto{4.6cm} having kernel rings of similar heights \\
\indent \indent Order \textsf{Echiniformes} \\
\indent \indent \indent Family \textsf{Echinidae (E)} \tabto{4.6cm} ``spiny bugs'', throny or wavy species \\
\indent \indent \indent Family \textsf{Geminidae (G)} \tabto{4.6cm} ``twin bugs'', two or more compartments \\
\indent \indent \indent Family \textsf{Ctenidae (Ct)} \tabto{4.6cm} ``comb bugs'', \textsf{P} with narrow strips \\
\indent \indent \indent Family \textsf{Uridae (U)} \tabto{4.6cm} ``tailed bugs'', with tails of various lengths \\
\indent  Class \textsf{Endokernel} \tabto{4.6cm} having strong inner kernel rings \\
\indent \indent Order \textsf{Kroniformes} \\
\indent \indent \indent Family \textsf{Kronidae (K)} \tabto{4.6cm} ``crown bugs'', complex versions of \textsf{S, P} \\
\indent \indent \indent Family \textsf{Quadridae (Q)} \tabto{4.6cm} ``square bugs'', $4\times 4$ grids of masses \\
\indent \indent \indent Family \textsf{Volvidae (V)} \tabto{4.6cm} ``twisting bugs'', possibly complex \textsf{H} \\
\indent \indent Order \textsf{Radiiformes} \\
\indent \indent \indent Family \textsf{Dentidae (D)} \tabto{4.6cm} ``gear bugs'', rotating with gear-like units \\
\indent \indent \indent Family \textsf{Radiidae (R)} \tabto{4.6cm} ``radial bugs'', regular or star polygon shaped \\
\indent \indent \indent Family \textsf{Bullidae (B)} \tabto{4.6cm} ``bubble bugs'', bilateral with bubbles inside \\
\indent \indent \indent Family \textsf{Lapillidae (L)} \tabto{4.6cm} ``gem bugs'', radially distributed small rings \\
\indent \indent \indent Family \textsf{Folidae (F)} \tabto{4.6cm} ``petal bugs'', stationary with petal-like units \\
\indent \indent Order \textsf{Amoebiformes} \\
\indent \indent \indent Family \textsf{Amoebidae (A)} \tabto{4.6cm} ``amoeba bugs'', volatile shape and behavior \\
}

Much like real-world biology, the taxonomy of Lenia is tentative and is subject to revisions or redefinitions when more data is available.

\subsubsection{Naming}
Following Theo Jansen for naming artificial life using biological nomenclature (\textit{Animaris} spp.) \cite{Jansen2008}, each Lenia species was given a binomial name that describes its geometric shape (genus name) and behavior (species name) to facilitate analysis and communication. Alphanumeric code was given in the form ``\textit{BGUs}'' with initials of genus or family name (\textit{G}) and species name (\textit{s}), number of units (\textit{U}), and rank (\textit{B}).

Suffix ``-\textit{ium}'' in genus names is reminiscent of a bacterium or chemical elements, while suffixes ``-\textsf{inae}'' (subfamily), ``-\textsf{idae}'' (family), and ``-\textsf{iformes}'' (order) were borrowed from actual animal taxa.  Numeric prefix \footnote{Prefixes used are: \textit{Di-, Tri-, Tetra-, Penta-, Hexa-, Hepta-, Octa-, Nona-, Deca-, Undeca-, Dodeca-, Trideca-}, etc.} in genus names indicates the number of units, similar to organic compounds and elements (IUPAC names)

\subsection{Ecology}

We describe the parameter space of Lenia (``geography'') and the distribution of lifeforms (``ecology'').

\begin{figure}[!tb]
\centering
\makebox[\textwidth][c]{\includegraphics[width=1.2\textwidth,trim=4 4 4 4,clip]{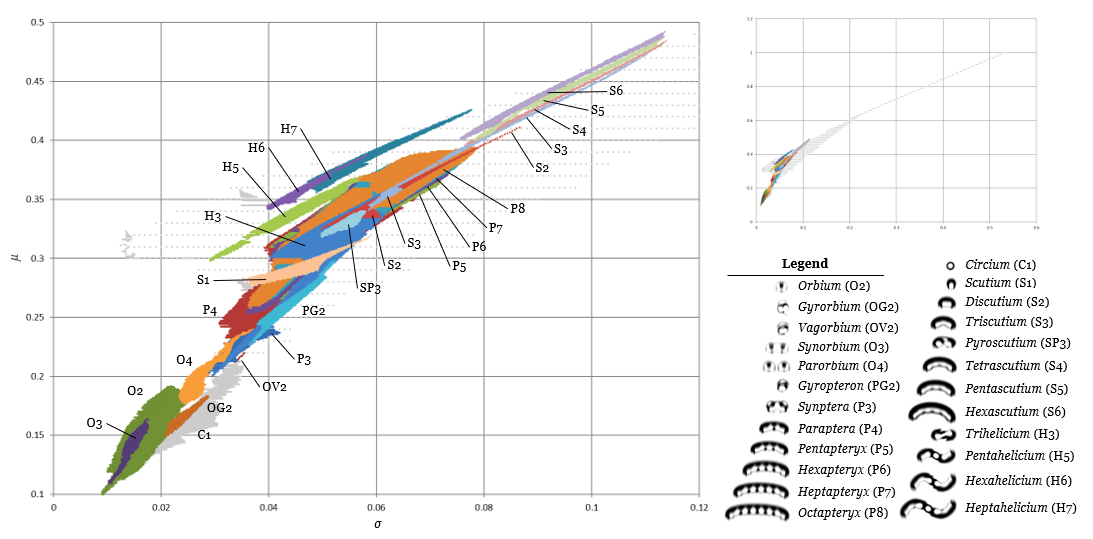}}
\caption{The $\mu$-$\sigma$ parameter space as $\mu$-$\sigma$ map, with niches of rank-1 species.  Total 142,338 loci.  \textbf{(legend)} Corresponding names and shapes for the species codes in the map.  \textbf{(inset)} Wider $\mu$-$\sigma$ map showing the niche of \textit{Circium} (grey region), demonstrates the four landscapes of rule space: class 1 homogenous desert (upper-left), class 2 cyclic savannah (central grey), class 3 chaotic forest (lower-right), class 4 complex river (central colored).}
\label{fig-map}
\end{figure}

\begin{figure}[!tb]
\centering
\includegraphics[width=\textwidth,trim=4 4 4 4,clip]{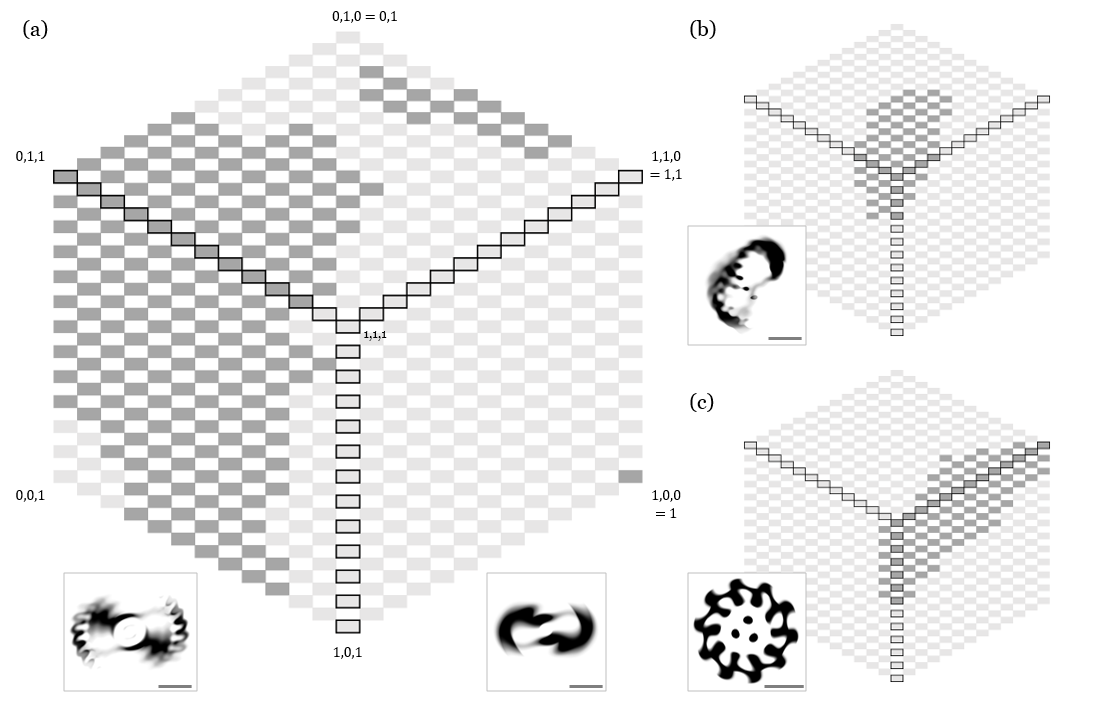}
\caption{The $\beta$ parameter space as $\beta$-cubes, with niches of selected species from the three Lenia classes.  \textbf{(a)} $\beta$-cube of class \textsf{Exokernel} exemplified by \textit{Helicium}, including rank-1 (right inset) niche at corner $(1, 0, 0)$, rank-2 (left inset) niche at edge near $(\frac{1}{2}, 1, 0)$, rank-3 niche on surfaces near $(\frac{1}{2}, \frac{1}{2}, 1)$.  \textbf{(b)} $\beta$-cube of class \textsf{Mesokernel} exemplified by \textit{Gyrogeminium gyrans} (inset), niche around $(1, 1, 1)$.  \textbf{(c)} $\beta$-cube of class \textsf{Endokernel} exemplified by \textit{Decadentium rotans} (inset), niche mostly on surface $(1, \beta_2, \beta_3$).}
\label{fig-cube}
\end{figure}

\subsubsection{Landscapes}
The four classes of CA rules \cite{Ilachinski2001, Wolfram2002} corresponds to the four \textit{landscapes} in the Lenia parameter space (Figure \ref{fig-map}):

\begin{itemize}[noitemsep]
\item Class 1 (homogenous ``desert'') produces no global or local pattern but a homogeneous (empty) state
\item Class 2 (cyclic ``savannah'') produces regional, periodic immobile patterns (e.g. \textit{Circium})
\item Class 3 (chaotic ``forest'') produces chaotic, aperiodic global filament network (``vegetation'')
\item Class 4 (complex ``river'') generates localized complex structures (lifeforms)
\end{itemize}

\subsubsection{Niches}
In the $(B+1)$-dimensional $\mu$-$\sigma$-$\beta$ parameter hyperspace, a lifeform only exists for a continuous parameter range called its \textit{niche}.  Each combination of parameters is called a \textit{locus} (plural: loci).

For a given $\beta$, a \textit{$\mu$-$\sigma$ map} is created by plotting the niches of selected lifeforms on a $\mu$ vs. $\sigma$ chart.  Maps of rank-1 species have been extensively charted and were used in taxonomical analysis (Figure \ref{fig-map}).

A \textit{$\beta$-cube} is created by marking the existence (or the size of $\mu$-$\sigma$ niche) of a lifeform at every $\beta$ locus.  As noted in ``Definition'' section, a $B$-dimensional hypercube can be reduced to its $(B-1)$-dimensional hypersurfaces, perfect for visualization in the three-dimensional case (Figure \ref{fig-cube}).

\subsection{Morphology}

We present the study of structural characteristics, or ``morphology'', of Lenia lifeforms.  See Figure \ref{fig-tax} for the family codes (\textsf{O, S, P}, etc.).

\begin{figure}[p]
\centering
\includegraphics[width=\textwidth,trim=4 4 4 4,clip]{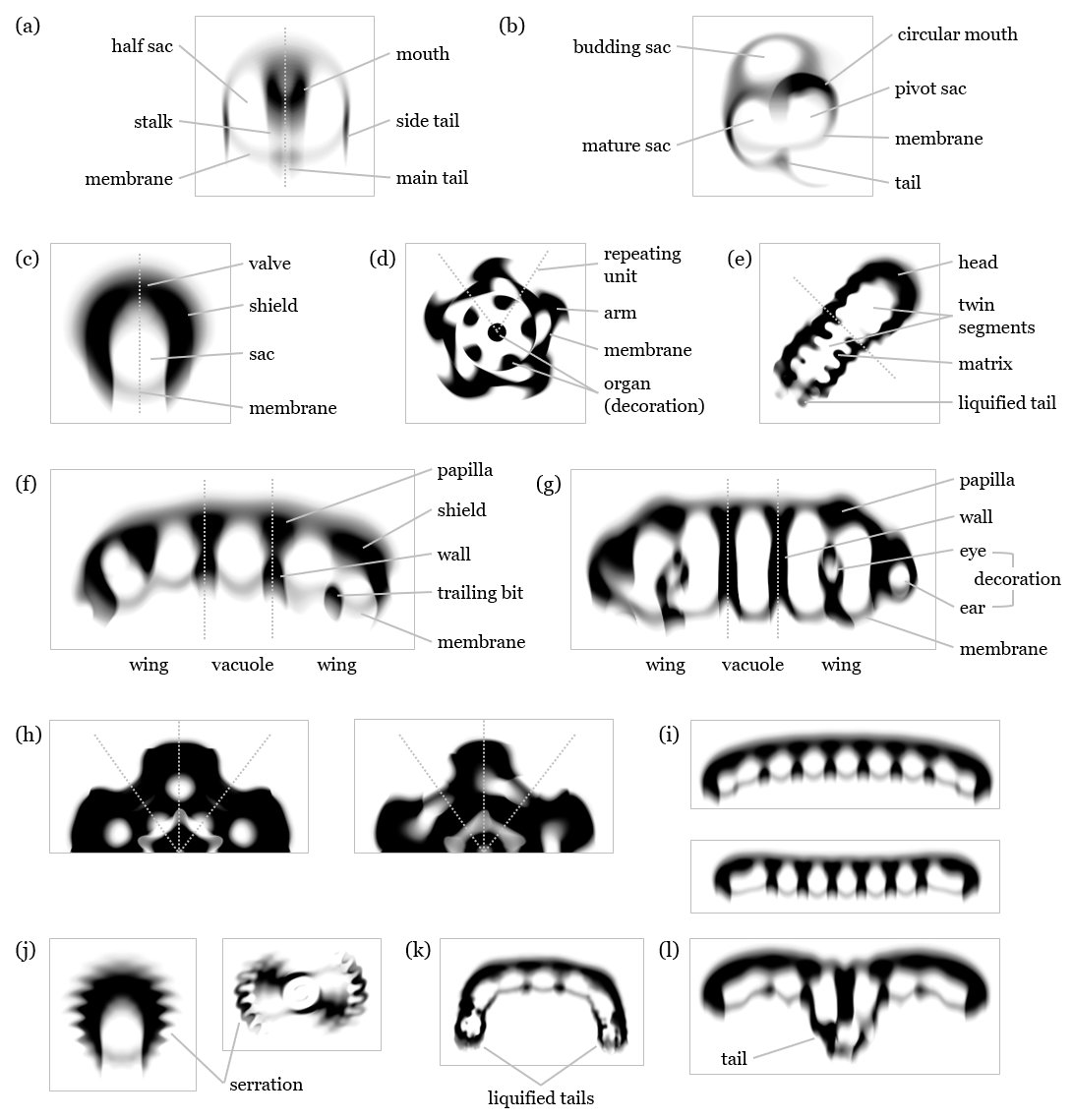}
\caption{Anatomy and symmetries in Lenia lifeforms (not to scale).  \textbf{(a-c)} Simple species as standalone components: \textit{Orbium} as standalone orb (a); \textit{Gyrorbium} as standalone orboid wing (b); \textit{Scutium} as standalone scutum (c).  \textbf{(d-g)} Complex species: radial \textit{Asterium rotans} (d); roughly bilateral \textit{Hydrogeminium natans} (e); long-chain \textit{Pentapteryx} (f) and \textit{Pentakronium} (g).  \textbf{(h)} Symmetry of radial units: bilateral units in stationary \textit{Asterium inversus} (left) and asymmetric units in rotational \textit{A. torquens} (right).  \textbf{(i)} Convexity: convex \textit{Nonapteryx arcus} (top) and concave \textit{N. cavus} (bottom).  \textbf{(j-l)} Ornamentation: serration in higher-rank \textit{Scutium} and \textit{Helicium} (j); liquefaction in \textit{Heptageminium natans} (k), also (e); caudation in \textit{Octacaudopteryx} (l).}
\label{fig-ana}
\end{figure}

\subsubsection{Architecture}
Lenia lifeforms possess morphological structures of various kinds, but they can be summarized into the following types of architectures:

\begin{itemize}[noitemsep]
\item \textit{Segmented architecture} is the serial combination of a few basic components, prevalent in class \textsf{Exokernel} (\textsf{O, S, P, H}), also \textsf{Ct, U, K}.
\item \textit{Radial architecture} is the radial arrangement of repeating units, common in \textsf{Radiiformes} in class \textsf{Endokernel} (\textsf{D, R, B, L, F}), also \textsf{C, E, V}.
\item \textit{Swarm architecture} is the volatile cluster of granular masses, not confined to a particular geometry or locomotion, as in \textsf{G, Q, A}.
\end{itemize}

\subsubsection{Components and Metamerism}
Segmented architecture is composed of the following inventory of components (class \textsf{Exokernel} only) (Figure \ref{fig-ana}(a-c, f)).

\begin{itemize}[noitemsep]
\item The \textit{orb} (disk) is a circular disk halved by a central stalk, found in \textsf{O}.
\item The \textit{scutum} (shield) is a disk with a thick front shield, found in \textsf{S}.
\item The \textit{wing} has two versions: the \textit{orboid} (disk-like) wing is a distorted orb with a budding mechanism that creates and destroys sacs repeatedly, found in concave \textsf{S, P, H}; the \textit{scutoid} (shield-like) wing is a distorted scutum, found in convex \textsf{S, P, H}.
\item The \textit{vacuole} (sac) is a disk between the wings of long-chain \textsf{S, P, H}.
\end{itemize}

Many of these components are possibly interrelated, e.g. the orboid wing and the orb, the scutoid wing and the scutum, as suggested by the similarity or smooth transitions between species.

Multiple components can be combined serially into long-chains through fusion or adhesion (e.g. Figure \ref{fig-tax} (O:2) or (O:1)), in a fashion comparable to \textit{metamerism} in biology (or \textit{multicellularity} if we consider the components as ``cells'') (Figure \ref{fig-ana}(f-g)).

Long-chain species exhibit different degrees of \textit{convexity}, from convex to concave: \textsf{S} $>$ convex \textsf{P} (\textit{arcus} subgenus) $>$ linear \textsf{O} $>$ concave \textsf{P} (\textit{cavus} subgenus); sinusoidal \textsf{P} (\textit{sinus} subgenus) have hybrid convexity (Figure \ref{fig-ana}(i), \ref{fig-tax} column 1).

Higher-rank segmented \textsf{Ct, U, K} also exhibit metamerism and convexity with more complicated components.

\subsubsection{Symmetry and Asymmetry}
\textit{Structural symmetry} is a prominent characteristic of Lenia life, including the following types:

\begin{itemize}[noitemsep]
\item \textit{Bilateral symmetry} (dihedral group D\textsubscript{1}) mostly in segmented and swarm architectures (\textsf{O, S, P, Ct, U, K; G, Q}).
\item \textit{Radial symmetry} (dihedral group D\textsubscript{n}) is geometrically rotational plus reflectional symmetry, caused by bilateral repeating units in radial architecture (\textsf{R, L, F, E}).
\item \textit{Rotational symmetry} (cyclic group C\textsubscript{n}) is geometrically rotational without reflectional symmetry, caused by asymmetric repeating units in radial architecture (\textsf{D, R, L}) (Figure \ref{fig-ana}(h)).
\item \textit{Spherical symmetry} (orthogonal group O(2)) is a special case of radial symmetry (\textsf{C}).
\item Secondary symmetries:
\begin{itemize}[noitemsep]
\item \textit{Spiral symmetry} is secondary rotational symmetry derived from twisted bilaterals (\textsf{H, V}).
\item \textit{Biradial symmetry} is secondary bilateral symmetry derived from radials (\textsf{B, R, E}).
\item \textit{Deformed bilateral symmetry} is bilateral with heavy asymmetry (e.g. gyrating species in \textsf{O, S, G, Q}).
\end{itemize}
\item No symmetry in amorphous species (\textsf{A}).
\end{itemize}

\textit{Asymmetry} also plays a significant role in shaping the lifeforms and guiding their movements, causing various degrees of angular motions (detailed in ``Physiology'' section).  Asymmetry is usually intrinsic in a species, as demonstrated by experiments where a slightly asymmetric form (e.g. \textit{Paraptera pedes}, \textit{Echinium limus}) was mirrored into perfect symmetry and remained metastable, but after the slightest perturbation (e.g. rotate 1\degree), it slowly restores to its natural asymmetric form.

\subsubsection{Ornamentation}
Many detailed local patterns arise in higher-rank species owing to their complex kernels (Figure \ref{fig-ana}(d-e, j-l)):

\begin{itemize}[noitemsep]
\item \textit{Decoration} is the addition of tiny ornaments (e.g. dots, circles, crosses), prevalent in class \textsf{Endokernel}.
\item \textit{Serration} is a ripple-like sinusoidal boundary or pattern, common in class \textsf{Exokernel} and \textsf{Mesokernel}.
\item \textit{Caudation} is a tail-like structure behind a long-chain lifeform (e.g. \textsf{P, K, U}), akin to ``tag-along'' in GoL.
\item \textit{Liquefaction} is the degradation of an otherwise regular structure into a chaotic ``liquified'' tail.
\end{itemize}

\subsection{Behavior}

We present the study of behavioral dynamics of Lenia lifeforms, or ``ethology'', in analogy to the study of animal behaviors in biology.
 
\begin{figure}[p]
\centering
\includegraphics[width=\textwidth,trim=4 4 4 4,clip]{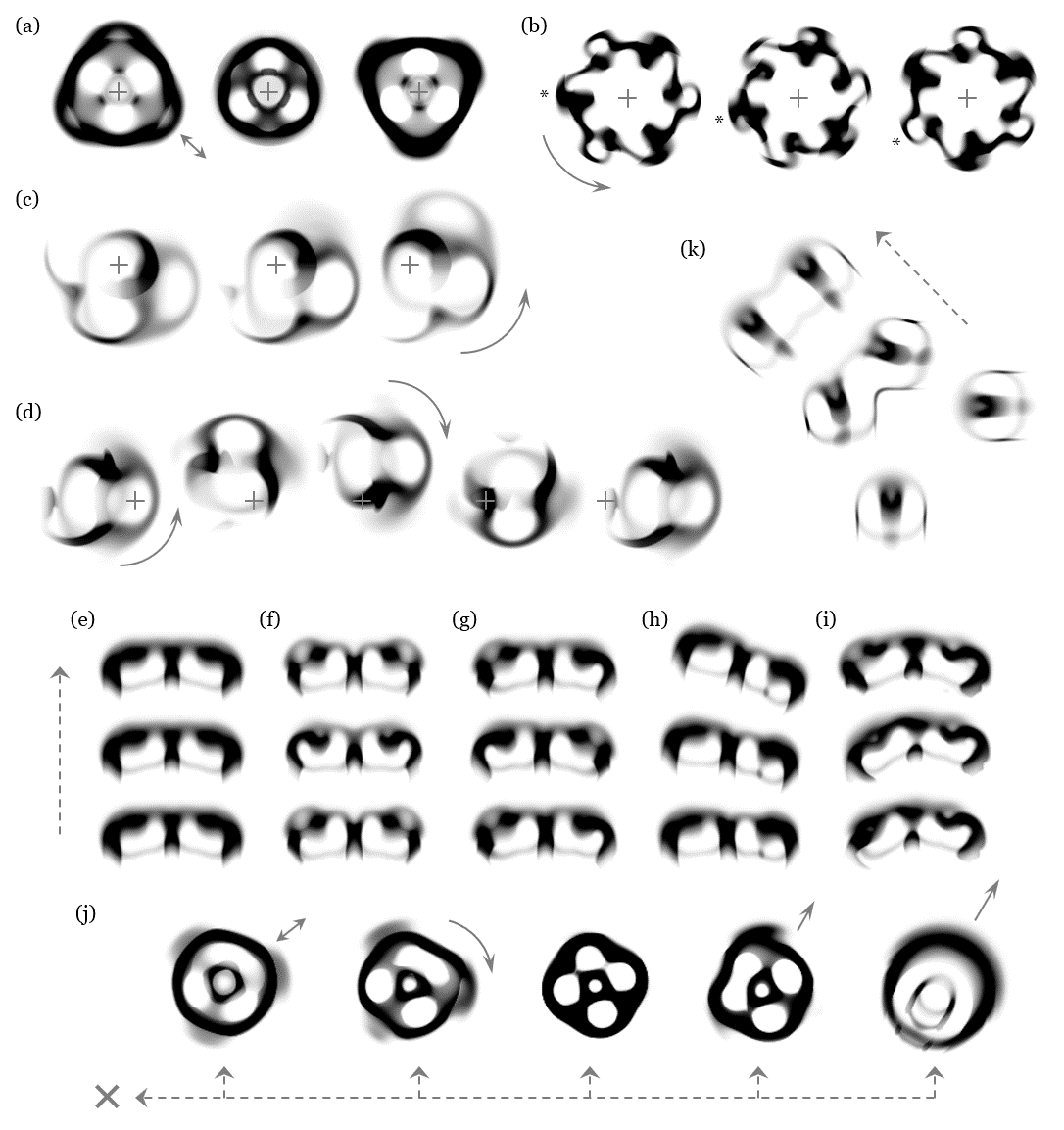}
\caption{Behavioral dynamics in Lenia lifeforms. (not to scale; $+$ = reference point; 
$\rightarrow$ = motion; $\dashrightarrow$ = time flow, left to right if unspecified) \textbf{(a)} Stationarity: inverting \textit{Trilapillium inversus} (S\textsubscript{O}).  \textbf{(b)} Rotation: twinkling \textit{Hexadentium scintillans} (R\textsubscript{A}) ($*$ = same unit).  \textbf{(c-d)} Gyration: gyrating \textit{Gyrorbium gyrans} (G\textsubscript{O}) (c); zigzagging \textit{Vagorbium undulatus} (G\textsubscript{A}) (d).  \textbf{(e-i)} Translocation with various gaits: sliding \textit{Paraptera cavus labens} (T\textsubscript{F}) (e); jumping \textit{P. c. saliens} (T\textsubscript{O}) (f); walking \textit{P. c. pedes} (T\textsubscript{A}) (g); deflected \textit{P. sinus pedes} (T\textsubscript{DA}) (h); chaotic \textit{P. s. p. rupturus} (T\textsubscript{CDA}) (i).  \textbf{(j)} Spontaneous metamorphosis: \textit{Tetralapillium metamorpha} switching among oscillating (S\textsubscript{A}), rotating (R\textsubscript{O}), frozen (S\textsubscript{F}), walking (T\textsubscript{A}), and wandering (T\textsubscript{C}) (left to right), occasionally die out ($\times$).  \textbf{(k)} Particle reactions: two \textit{Orbium} collide and fuse together into an intermediate, then stabilize into one \textit{Synorbium}.}
\label{fig-beh}
\end{figure}

\subsubsection{Locomotion}
In GoL, pattern behaviors include stationary (fixed, oscillation), directional (orthogonal, diagonal, rarely oblique), and infinite growth (linear, sawtooth, quadratic) \cite{LifeWiki}.  SmoothLife added omnidirectional movement to the list \cite{Rafler2011}.  Lenia supports a qualitatively different repertoire of behaviors, which can be described in global and local levels.

The global movements of lifeforms are summarized into \textit{modes of locomotion} (Figure \ref{fig-beh}(a-c, e)):

\begin{itemize}[noitemsep]
\item \textit{Stationarity} (S) means the pattern stays still with negligible directional movement or rotation.
\item \textit{Rotation} (R) is the angular movement around a stationary centroid.
\item \textit{Translocation} (T) is the directional movement in certain direction.
\item \textit{Gyration} (G) is the angular movement around a non-centroid center, basically a combination of translocation and rotation.
\end{itemize}

In formula,
\begin{equation}
\mathbf{A}^{t+\tau} \approx (S_{s\tau} \circ R_{\omega\tau})(\mathbf{A}^t)
\quad \begin{cases}
\text{Stationarity:}& s = 0, \omega = 0 \\
\text{Rotation:}& s = 0, \omega > 0 \\
\text{Translocation:} & s > 0, \omega = 0 \\
\text{Gyration:}& s > 0, \omega > 0 \\
\end{cases}
\end{equation}
where $\tau$ is the quasi-period, $S$ is a shift by distance $s\tau$ due to linear speed $s$, $R$ is a rotation (around the centroid) by angle $\omega\tau$ due to angular speed $\omega$.

\subsubsection{Gaits}
The local details of movements are identified as different \textit{gaits} (Figure \ref{fig-beh}(e-i)):

\begin{itemize}[noitemsep]
\item \textit{Fixation} (\textsubscript{F}) means negligible or no fluctuation during locomotion.
\item \textit{Oscillation} (\textsubscript{O}) is the periodic fluctuation during locomotion.
\item \textit{Alternation} (\textsubscript{A}) is global oscillation plus out-of-phase local oscillations (see ``Physiology'' section).
\item \textit{Deviation} (\textsubscript{D}) is a small departure from the regular locomotion, e.g. slightly curved linear movement, slight movements in the rotating or gyrating center.
\item \textit{Chaoticity} (\textsubscript{C}) is the chaotic, aperiodic movements.
\end{itemize}

Any gait or gait combination can be coupled with any locomotive mode, and is represented by the combined code (e.g. chaotic deviated alternating translocation = T\textsubscript{CDA}).  See Table \ref{table-sym} for all combinations.

\begin{table}
\centerline{\scriptsize\begin{tabular}{|l|llll|}
\hline
Gait &\multicolumn{4}{|c|}{Locomotive mode (type of symmetry)} \\
\cline{2-5}
(type of &Stationarity &Rotation &Translocation &Gyration \\
\  asymmetry)&(Radial) &(Rotational) &(Bilateral) &(Deformed \\
& & & &\  bilateral) \\
\hline
Fixation &S\textsubscript{F} &R\textsubscript{F} &T\textsubscript{F} &G\textsubscript{F} \\
(Static) &= Frozen &= Rotating &= Sliding &= Spinning \\
&\textit{Pentafolium} &\textit{Asterium} &\textit{Paraptera} &\textit{Gyropteron} \\
&\textit{lithos} &\textit{rotans} &\textit{cavus labens} [e] &\textit{serratus velox} \\
\hline
Oscillation &S\textsubscript{O} &R\textsubscript{O} &T\textsubscript{O} &G\textsubscript{O} \\
(Dynamical) &= Ventilating &= Torqueing &= Jumping &= Gyrating \\
&\textit{Hexalapillium} &\textit{Asterium} &\textit{Paraptera} &\textit{Gyrorbium} \\
&\textit{ventilans} &\textit{torquens} &\textit{cavus saliens} [f] &\textit{gyrans} [c] \\
\hline
Alternation &S\textsubscript{A} &R\textsubscript{A} &T\textsubscript{A} &G\textsubscript{A} \\
(Out-of-phase) &= Inverting &= Twinkling &= Walking &= Zigzagging \\
&\textit{Trilapillium} &\textit{Hexadentium} &\textit{Paraptera} &\textit{Vagorbium} \\
&\textit{inversus} [a] &\textit{scintillans} [b] &\textit{cavus pedes} [g] &\textit{undulatus} [d] \\
\hline
Deviation &S\textsubscript{D} &R\textsubscript{D} &T\textsubscript{D} &G\textsubscript{D}  \\
(Unbalanced) &= Drifting &= Precessing &= Deflected &= Revolving \\
&\textit{Octafolium} &\textit{Nivium} &\textit{Paraptera} &\textit{Gyrorbium} \\
&\textit{tardus} &\textit{incarceratus} &\textit{sinus pedes} [h] &\textit{revolvens} \\
\hline
Chaoticity &S\textsubscript{C} &R\textsubscript{C} &T\textsubscript{C} &G\textsubscript{C} \\
(Stochastic) &= Vibrating &= Tumbling &= Wandering &= Swirling \\
&\textit{Asterium} &\textit{Decadentium} &\textit{Paraptera} &\textit{Gyrogeminium} \\
&\textit{nausia} &\textit{volubilis} &\textit{s. p. rupturus} [i] &\textit{velox} \\
\hline
\end{tabular}}
\caption{Matrix of symmetries, asymmetries, locomotive modes, and gaits.  Each combination is provided with a code, a descriptive term and a sample species.  (Brackets indicate sub-figures in Figure \ref{fig-beh})}
\label{table-sym}
\end{table}

\subsubsection{Metamorphosis}
\textit{Spontaneous metamorphosis} is a highly chaotic behavior in Lenia, where a ``shapeshifting'' species frequently switch among different morphological-behavioral \textit{templates}, forming a continuous-time Markov chain.  Each template often resembles an existing species.  The set of possible templates and the transition probabilities matrix are determined by the species and parameter values (Figure \ref{fig-beh}(j)).

An extreme form of spontaneous metamorphosis is exhibited by the \textsf{Amoebidae}, where the structure and locomotive patterns are no longer recognizable, while a bounded size is still maintained.

These stochastic behaviors denied the previous assumption that morphologies and behaviors are fixed qualities in a species, but are actually probabilistic (albeit usually single template with probability one).

\subsubsection{Infinite Growth}
Unlike the above behaviors where the total mass remains finite, there are behaviors associated with infinite growth (positive or negative).

\textit{Explosion} or \textit{evaporation} is the uncontrolled infinite growth, where the mass quickly expands or shrinks in all directions, the lifeform fails to self-regulate and dies out.

\textit{Elongation} or \textit{contraction} is the controllable infinite growth, where a long-chain lifeform keeps lengthening or shortening in directions tangential to local segments.  Microscopically, vacuoles are being constantly created or absorbed via binary fission or fusion.

As estimated by mass time-series, linear and circular elongation show linear growth rate, while spiral elongation (in \textsf{Helicidae}) and others show quadratic growth rate.

\subsubsection{Particle Reactions}
Using the interactive program as a ``particle collider'' (cf. \cite{Martinez2017}), we investigated the reactions among \textsf{Orbidae} instances acting as physical or chemical particles.  They often exhibit elasticity and resilience during collision, engage in inelastic (sticky) collision, and seem to exert a kind of weak ``attractive force'' when two particles are nearby or ``repulsive force'' when getting too close.

Reaction of two or more \textit{Orbium} particles with different starting positions and incident angles would result in one of the followings:

\begin{itemize}[noitemsep]
\item \textit{Deflection}, two \textit{Orbium} disperse in different angles.
\item \textit{Reflection}, one \textit{Orbium} unchanged and one goes in opposite direction.
\item \textit{Absorption}, only one \textit{Orbium} survives.
\item \textit{Annihilation}, both \textit{Orbium} evaporates.
\item \textit{Detonation}, the resultant mass explodes into infinite growth.
\item \textit{Fusion}, multiple \textit{Orbium} fuse together into \textsf{Synorbinae} (Figure \ref{fig-beh}(k)).
\item \textit{Parallelism}, multiple \textit{Orbium} travel in parallel with ``forces'' subtly balanced, forming \textsf{Parorbinae} (Figure \ref{fig-tax} (O:1)).
\end{itemize}

Starting from a composite \textsf{Orbidae} may result in:

\begin{itemize}[noitemsep]
\item \textit{Fission}, one \textsf{Synorbinae} breaks into multiple \textsf{Synorbinae} or \textit{Orbium}.
\end{itemize}

\subsection{Physiology}

The exact mechanisms of morphogenesis (self-organization) and homeostasis (self-regulation), or ``physiology'', in Lenia are not well understood.  Here we will present a few observations and speculations.

\subsubsection{Symmetries and Behaviors}
A striking result in analyzing Lenia is the correlations between structural symmetries/asymmetries (``Morphology'' section) and behavioral dynamics (``Behavior'' section).

At a global scale, the locomotive modes (stationarity, rotation, translocation, gyration) correspond to the types of overall symmetry (radial, rotational, bilateral, deformed bilateral).  At a local scale, the locomotive gaits (fixation, oscillation, alternation, deviation, chaoticity) correspond to the development and distribution of asymmetry (static, dynamic, out-of-phase among units, unevenly distributed, stochastic development) (Table \ref{table-sym}).

\subsubsection{Stability-Motility Hypothesis}
A closer look in these symmetry-behavior correlations suggests the mechanisms of how motions arise.

In a bilateral species, while there is lateral (left-right) reflectional symmetry, the heavy asymmetry along the longitudinal (rostro-caudal) axis may be the origin of directional movement.  In a deformed bilateral species, the lateral symmetry is broken, thus introduces an angular component to its linear motion.

In a radial species, bilateral repeating units are arranged radially, all directional vectors cancel out, thus overall remain stationary.  In a rotational species, asymmetric repeating units mean the lateral symmetry is broken, thus initiates angular rotation around the centroid.

On top of these global movements, the dynamical qualities of asymmetry --- static/dynamical, in-phase/out-of-phase, balanced/unbalanced, regular/stochastic --- lead to the dynamical qualities of locomotion (i.e. gaits).

Based on these reasonings, we propose the \textit{stability-motility hypothesis} (potentially applicable to real-world physiology or evolutionary biology):
\begin{quote}
Symmetry provides stability; asymmetry provides motility. \\
\\
Distribution of asymmetry determines locomotive mode; its development determines gait.
\end{quote}

\subsubsection{Alternation and Internal Communication}
The alternation gait, that is global oscillation plus out-of-phase local oscillations, is one of the most complicated behavior in Lenia.  It demonstrates phenomena like long-range synchronization and rotational clockwork.

\begin{figure}[!tb]
\centering
\includegraphics[width=\textwidth,trim=4 4 4 4,clip]{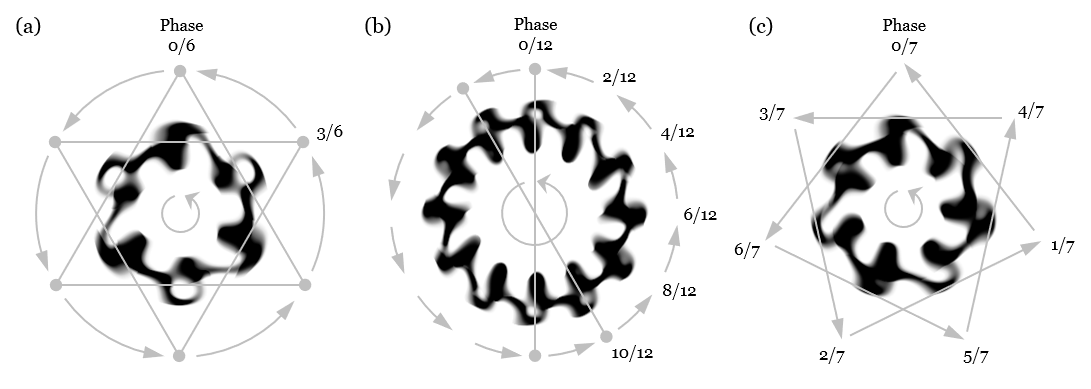}
\caption{``Rotational clockwork'' in selected alternating \textsf{Dentidae} species.  After $1/n$ cycle, all phases advance by $1/n$ while phase relations remain unchanged.  (not to scale; $\rightarrow$ = phase transfer; $\multimap$ = same phase; $\circlearrowleft$ = rotation, taken as the positive direction) \textbf{(a)} Even-sided \textit{Hexadentium scintillans}, with opposite-phase adjacent units and same-phase alternating units.  \textbf{(b)} Even-sided \textit{Dodecadentium scintillans}, with sequentially out-of-phase adjacent units and same-phase opposite units.  \textbf{(c)} Odd-sided \textit{Heptadentium scintillans}, with globalized phase distribution.}
\label{fig-alt}
\end{figure}

\begin{table}
\centerline{\scriptsize\begin{tabular}{|l|cccc|}
\hline
Genus &Rank &Units &Phase difference &Rotational symmetry \\
(species \textit{scintillans}) &($B$) &($n$) &($k/n$ cycle) of &($m$ units = angle $m \cdot 2 \pi / n$)\\
&&&adjacent units &between $1/n$ cycle \\
\hline
\textit{Hexadentium} [a] &$2$ &$6$ &$3 / 6$ &$1 \cdot 2\pi/6$ (adjacent) \\
\textit{Heptadentium} [c] &$2$ &$7$ &$4 / 7$ &$2 \cdot 2\pi/7$ (skipping) \\
\textit{Octadentium} &$2$ &$8$ &$4 / 8$ &$1 \cdot 2\pi/8$ (adjacent) \\
\textit{Nonadentium} &$2$ &$9$ &$5 / 9$ &$2 \cdot 2\pi/9$ (skipping) \\
\textit{Decadentium} &$4$ &$10$ &$2 / 10$ &$1 \cdot 2\pi/10$ (adjacent) \\
\textit{Undecadentium} &$4$ &$11$ &$2 / 11$ &$6 \cdot 2\pi/11$ (skipping) \\
\textit{Dodecadentium} [b] &$4$ &$12$ &$2 / 12$ &$1 \cdot 2\pi/12$ (adjacent) \\
\textit{Tridecadentium} &$4$ &$13$ &$3 / 13$ &$9 \cdot 2\pi/13$ (skipping) \\
\hline
\end{tabular}}
\caption{Alternation characteristics $(B, n, k, m)$ in selected alternating \textsf{Dentidae} species.  (Brackets indicate sub-figures in Figure \ref{fig-alt})}
\label{table-alt}
\end{table}

\textit{Alternating translocation} (T\textsubscript{A}) in a simple bilateral species, where the two halves are in opposite phases, is the spatiotemporal reflectional (i.e. glide) symmetry at half-cycle, in addition to the full oscillation:
\begin{align}
\mathbf{A}^{t+\frac{\tau}{2}} & \approx (S_{\frac{s \tau}{2}} \circ F) (\mathbf{A}^t) \\
\mathbf{A}^{t+\tau} & \approx S_{s\tau} (\mathbf{A}^t)
\end{align}
where $\tau$ is the quasi-period, $S$ is a shift, $F$ is a flip. (Figure \ref{fig-beh}(g))

Alternating long-chain species, where two wings are oscillating out-of-phase but the main chain remains static, demonstrates \textit{long-range synchronization} in which faraway structures are able to synchronize.\footnote{We performed experiments to show that alternation is self-recovering, meaning that it is not coincidental but actively maintained by the species.}

\textit{Alternating gyration} (G\textsubscript{A}) is a special case in \textit{Vagorbium} (a variant of \textit{Gyrorbium}) where it gyrates to the opposite direction every second cycle, resulting in a zig-zag trajectory (Figure \ref{fig-beh}(d)).

\textit{Alternating stationarity} (S\textsubscript{A}) occurs in stationary radial lifeforms (with $n$ repeating units), leads to spatiotemporal reflectional (or rotational) symmetry at half-cycle:
\begin{align}
\mathbf{A}^{t+\frac{\tau}{2}} & \approx F(\mathbf{A}^t) \approx R_{\frac{\pi}{n}}(\mathbf{A}^t) \\
\mathbf{A}^{t+\tau} & \approx \mathbf{A}^t
\end{align}
where $R$ is a rotation.  This gives an optical illusion of ``inverting'' motions (Figure \ref{fig-beh}(a)).

\textit{Alternating rotation} (R\textsubscript{A}) is an intricate phenomenon found in rotational species, especially family \textsf{Dentidae}.  Consider a \textsf{Dentidae} species with $n$ repeating units, two adjacent units are separated spatially by angle $2\pi / n$ and temporally by $k/n$ cycle, $k \in \mathbb{Z}$ (Figure \ref{fig-alt}).  After $1/n$ cycle, the pattern recreates itself with rotation due to angular speed $\omega$, plus an extra spatiotemporal rotational symmetry of $m$ units due to pattern alternation, $m \in \mathbb{Z}$:
\begin{align}
\mathbf{A}^{t+\frac{\tau}{n}} & \approx R_{\frac{\omega\tau}{n} + \frac{2 \pi m}{n}}(\mathbf{A}^t) \\
\mathbf{A}^{t+\tau} & \approx R_{\omega\tau} (\mathbf{A}^t)
\end{align}
This giving an illusion that local features (e.g. a hole) are transferring from one unit to another (Figure \ref{fig-alt} outer arrows).  The values of $k, m$ seem to follow some particular trend (Table \ref{table-alt}).

\subsubsection{Allometry}
Besides direct observation, Lenia patterns were studied through statistical measurement and analysis, akin to ``allometry'' in biology.

\begin{figure}[!tb]
\centering
\includegraphics[width=\textwidth,trim=4 4 4 4,clip]{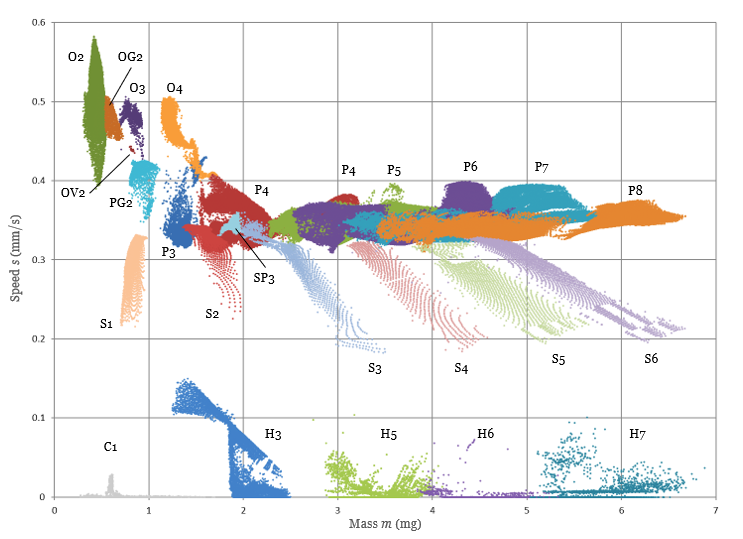}
\caption{Allometric chart of linear speed $s_m$ vs. mass $m$ for rank-1 species.  Total 142,338 loci, 300 time-steps ($t=30 \text{s}$) per locus.  See Figure \ref{fig-map} legend for species codes.}
\label{fig-mea}
\end{figure}

\begin{table}[!tb]
\centerline{\scriptsize\begin{tabular}{|l|ccc|}
\hline
Locomotion modes &Measure of &Measure of &Measure of \\
and gaits &linear motion &angular motion &oscillation \\
&($s_m$) &($|m_\Delta|, \omega_m, \omega_s, \ldots$) &($m, g, s_m, \ldots$) \\
\hline
Stationarity &Average $\approx$ 0 &Average $\approx$ 0 & \\
Rotation &Average $\approx$ 0 &\; Average $>$ 0 * & \\	
Translocation &Average $>$ 0 &Average $\approx$ 0 & \\
Gyration &Average $>$ 0 &Average $>$ 0 & \\
\hline
Fixation & & &Variability $\approx$ 0 \\
Oscillation & & &Variability $>$ 0 \\
Alternating translocation & &Variability $>$ 0 & \\
Deviated translocation & &\; Average $>$ 0 $\dagger$ & \\
Chaoticity &\multicolumn{3}{c|}{$\longleftarrow$ Chaotic trajectory $\longrightarrow$} \\
\hline
\end{tabular}}
\caption{Allometric relationships between behavior and statistical measures.  {\footnotesize (*~Only works in some cases; $\dagger$~Deviated translocation is similar to gyration)}}
\label{table-mea}
\end{table}

Various behaviors were found related to the average (mean or median), variability (standard deviation or interquartile length) or phase space trajectory of various statistical measures (Table \ref{table-mea}).

A few general trends were deduced from allometric charts, for example, linear speed is found to be roughly inverse proportional to density.  From the linear speed $s_m$ vs. mass $m$ chart (Figure \ref{fig-mea}), genera form strata according to linear speed (\textsf{O$>$P$>$S$>$H$>$C}), and species form clusters according to mass.

\subsection{Case Study}

In previous sections, we outlined the general characterizations of Lenia from various perspectives.  Here we combine these aspects in a focused study of one representative genus --- \textit{Paraptera} (P4) --- as a demonstration of concrete qualitative and quantitative analysis.

\subsubsection{The Unit-4 Group}
\textit{Paraptera} (P4) is closely related to two other genera \textit{Parorbium} (O4) and \textit{Tetrascutium} (S4), they comprise the rank-1 unit-4 group.

\begin{figure}[!tb]
\centering
\includegraphics[width=\textwidth,trim=4 4 4 4,clip]{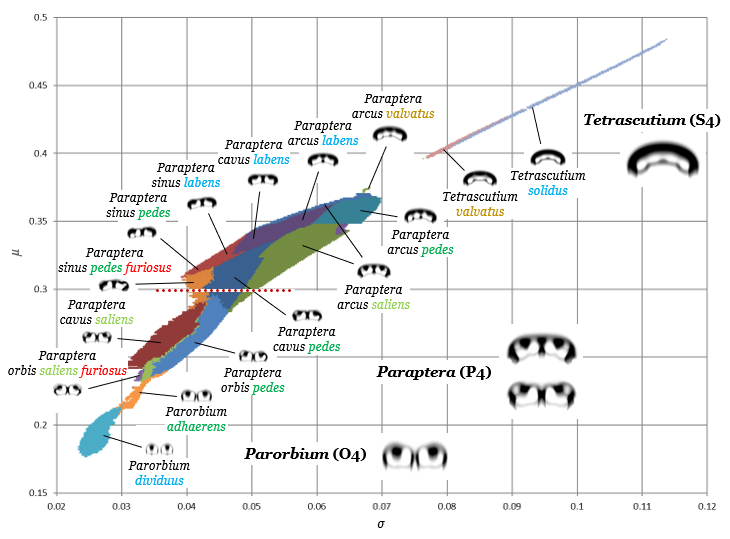}
\caption{$\mu$-$\sigma$ map of the unit-4 group, showing the prominent \textit{Parorbium-Paraptera-Tetrascutium} complex.  Total 16,011 loci.  The red dotted line marks the cross-sectional study (Figure \ref{fig-u4mea}).}
\label{fig-u4map}
\end{figure}

\begin{table}
\centerline{\scriptsize\begin{tabular}{|lll|ll|}
\hline
\multicolumn{3}{|l|}{Species} &Morphology &Behavior \\
\hline
O4 &\multicolumn{4}{l|}{genus \textit{Parorbium} (family \textsf{Orbidae})} \\
\hline
&O4d &\textit{Po. dividuus} &Two parallel orbs, separated &T = translocating \\
&O4a &\textit{Po. adhaerens} &Two parallel orbs, adhered &T \\
\hline
P4 &\multicolumn{4}{l|}{genus \textit{Paraptera} (family \textsf{Pterifera})} \\
\hline
&P4o* &\textit{P. orbis} * &Concave, twin orboid wings &T \\
&P4c* &\textit{P. cavus} * &Concave, twin orboid wings &T \\
&P4a* &\textit{P. arcus} * &Convex, twin scutoid wings &T \\
&P4s* &\textit{P. sinus} * &Sinusoidal, orboid + scutoid wings &T\textsubscript{D*} = deflected \\
&P4*l &\textit{P.} * \textit{labens} &Bilateral &T\textsubscript{F} = sliding \\
&P4*s &\textit{P.} * \textit{saliens} &Bilateral &T\textsubscript{O} = jumping \\
&P4*p &\textit{P.} * \textit{pedes} &Bilateral with slight asymmetry &T\textsubscript{A} = walking \\
&P4*v &\textit{P.} * \textit{valvatus} &\textsf{Scutidae}-like, twin wings, valving &T\textsubscript{O} = valving \\
&P4**f &\textit{P.} * * \textit{furiosus} &Occasional stretched wing &T\textsubscript{C*} = chaotic \\
\hline
S4 &\multicolumn{4}{l|}{genus \textit{Tetrascutium} (family \textsf{Scutidae})} \\
\hline
&S4s &\textit{T. solidus} &Four fused scuta, solid &T\textsubscript{F} = sliding \\
&S4v &\textit{T. valvatus} &Four fused scuta, valving &T\textsubscript{O} = valving \\
\hline
\end{tabular}}
\caption{Non-exhausive list of species identified in the unit-4 group.  {\footnotesize (* = Combinations are possible, e.g. P4spf with behavior T\textsubscript{CDA})}}
\label{table-u4}
\end{table}

\begin{figure}
\centering
\includegraphics[width=\textwidth,trim=4 4 4 4,clip]{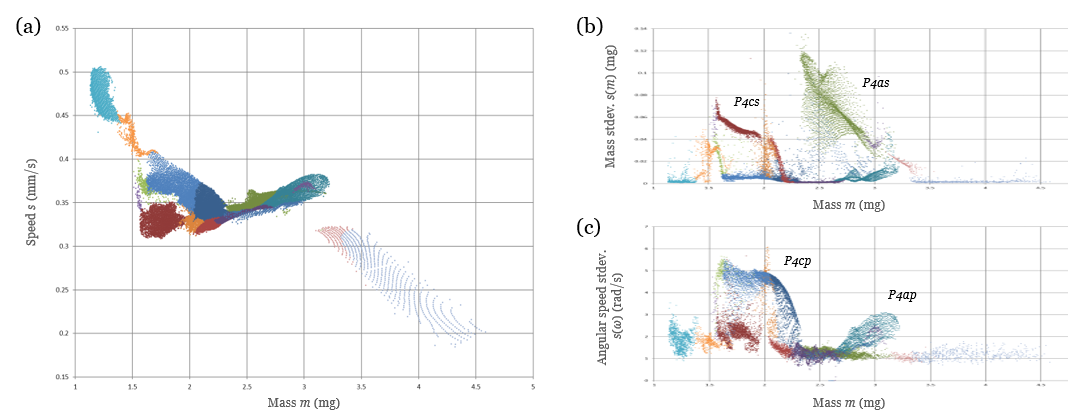}
\caption{Allometric charts of various measures for the unit-4 family.  Total 16,011 loci, 300 time-steps ($t=30 \mathrm{s}$) per locus.  \textbf{(a)} Linear speed $s_m$ vs. mass $m$, similar to the $\mu$-$\sigma$ map flipped.  \textbf{(b)} Mass variability $s(m)$ vs. mass $m$, isolates jumping (T\textsubscript{O}) species P4cs and P4as.  \textbf{(c)} Angular speed variability $s(\omega_m)$ vs. mass $m$, isolates walking (T\textsubscript{A}) species P4cp and P4ap.}
\label{fig-u4mea}
\end{figure}

In the $\mu$-$\sigma$ map (Figure \ref{fig-u4map}), their niches comprise the \textit{Parorbium-Paraptera-Tetrascutium} (O4-P4-S4) complex.  The narrow bridge between O4 and P4 indicates possible continuous transformation, and the agreement between the small tip of P4 and S4 suggests a remote relationship.  Species were isolated using allometric methods (Figure \ref{fig-u4mea}, Table \ref{table-mea}), verified in simulation, and assigned new names (Table \ref{table-u4}).

\subsubsection{Cross-Sectional Study}
\begin{figure}
\centering
\includegraphics[width=\textwidth,trim=4 4 4 4,clip]{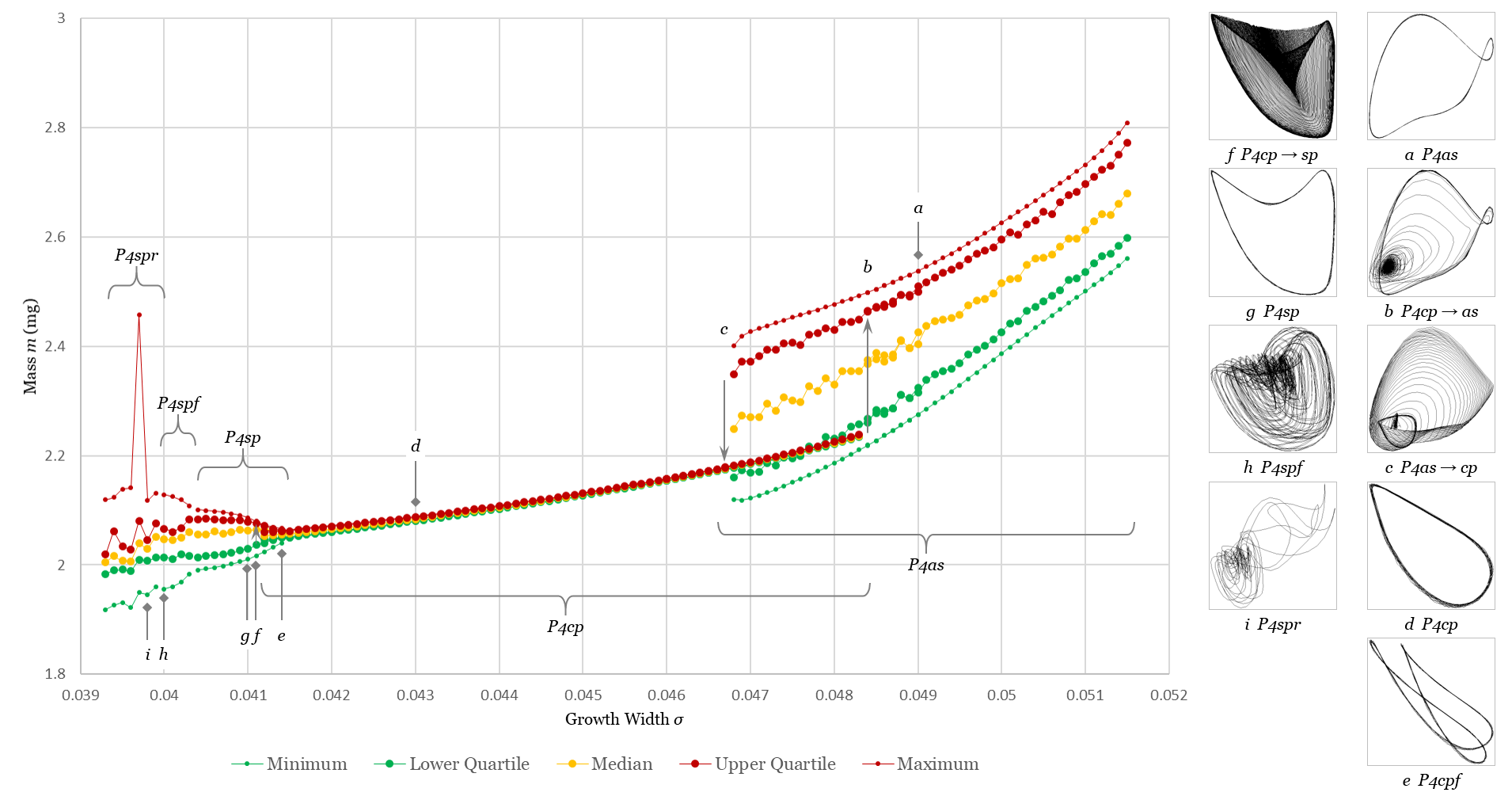}
(a)
\includegraphics[width=\textwidth,trim=4 4 4 4,clip]{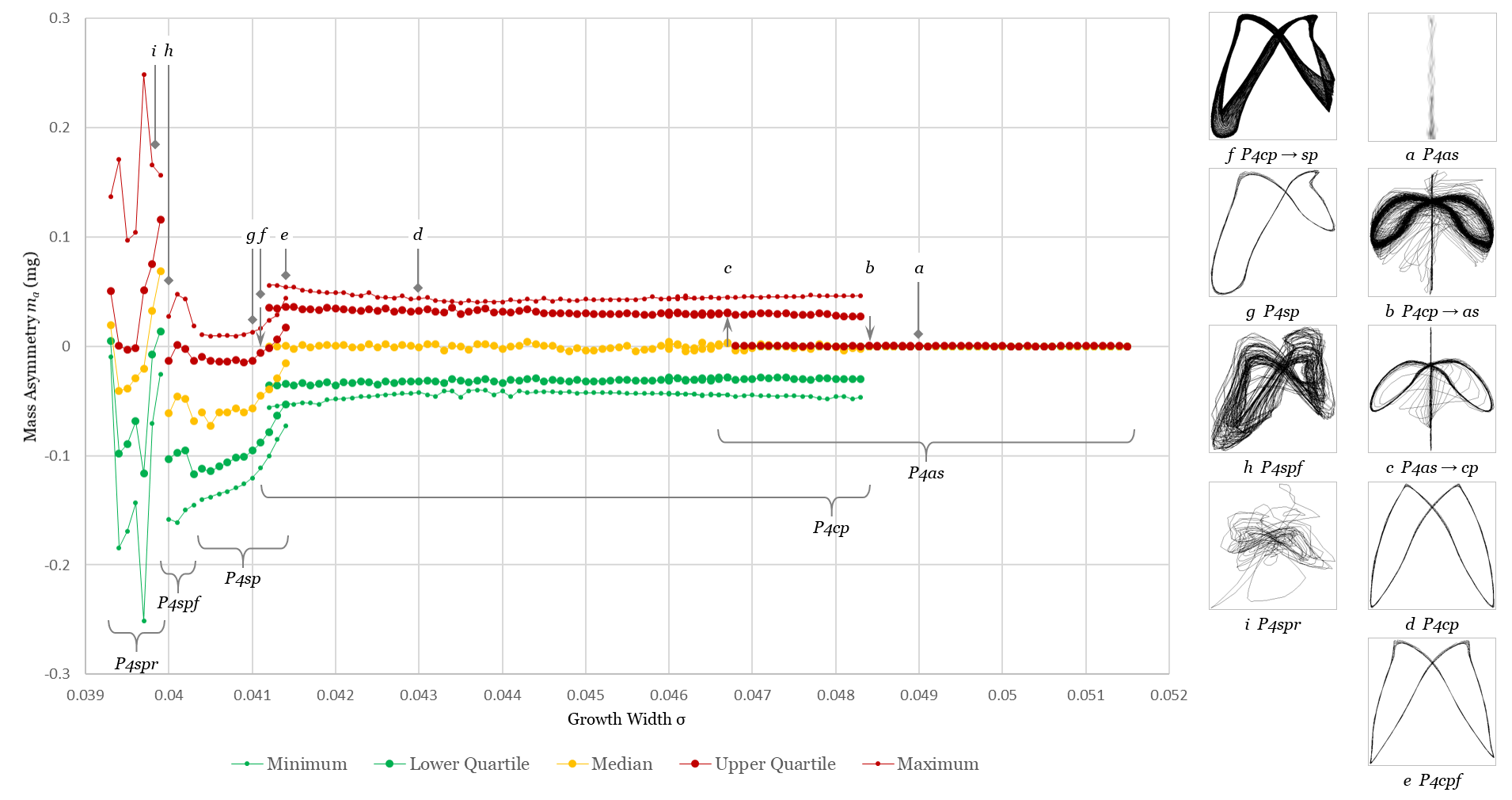}
(b)
\caption{Cross-sectional charts at $\mu=0.3, \sigma \in [0.0393, 0.0515]$ in genus \textit{Paraptera}. 200 time-steps ($t=20 \mathrm{s}$) per locus (See Table \ref{table-cross} for species codes).  \textbf{(a)} Mass $m$ vs. parameter $\sigma$, insets: growth $g$ vs. mass $m$ phase space trajectories at loci a-i.  \textbf{(b)} Mass asymmetry $m_\Delta$ vs. parameter $\sigma$, insets: linear speed $s_m$ vs. angular speed $\omega_m$ phase space trajectories at loci a-i.}
\label{fig-cross}
\end{figure}

In P4, a cross section at $\mu=0.3$ was further investigated, where five species exist in $\sigma \in [0.0393, 0.0515]$ (Figure \ref{fig-u4map} red dotted line, Table \ref{table-cross}).  Their behavioral traits were assessed via cross-sectional charts and snapshot phase space trajectories (Figure \ref{fig-cross}, see also Figure \ref{fig-beh}(e-i)).

\begin{table}
\centerline{\scriptsize\begin{tabular}{|lll|l|}
\hline
\multicolumn{2}{|l}{Species} &$\sigma$ range &Morphology and behavior \\
\hline
P4as &\textit{P. arcus saliens} &$[0.0468, 0.0515]$ &Convex, jumping (T\textsubscript{O}) \\
P4cp &\textit{P. cavus pedes} &$[0.0412, 0.0483]$ &Concave, walking (T\textsubscript{A}) \\
P4sp &\textit{P. sinus pedes} &$[0.0404, 0.0414]$ &Sinusoidal, deflected walking (T\textsubscript{DA}) \\
P4spf &\textit{P. s. p. furiosus} &$[0.0400, 0.0403]$ &As above plus chaotic (T\textsubscript{CDA}) \\
P4spr &\textit{P. s. p. rupturus} &$[0.0393, 0.0399]$ &As above plus fragile \\
\hline
\end{tabular}}
\caption{List of \textit{Paraptera} species in the cross-section $\mu=0.3$.}
\label{table-cross}
\end{table}

At higher $\sigma$ values, \textit{Paraptera arcus saliens} (P4as) has high $m$ variability and near zero $m_\Delta$, corresponding to their jumping behavior and perfect bilateral symmetry (locus a).  \textit{P. cavus pedes} (P4cp) has high $m_\Delta$ variability, matching their walking behavior and alternating asymmetry (locus d).

Just outside the coexistence of P4as and P4cp over $\sigma \in [0.0468, 0.0483]$, they slowly transform into each other, as shown by the spiral phase space trajectories (loci b, c).  Similarly for P4cp and P4sp (locus f).

Irregularity and chaos arise at lower $\sigma$.  For \textit{P. sinus pedes} (P4sp), non-zero $m_\Delta$ indicates deflected movement and asymmetry (locus g).  For \textit{P. sinus pedes furiosus} (P4spf), chaotic phase space trajectory indicates to chaotic movement and deformation (locus h).

At the edge of chaoticity, \textit{P. sinus pedes rupturus} (P4spr) has even higher and rugged variability, often encounters episodes of acute deformation but eventually recovers (locus i).  Outside the $\sigma$ lower bound, the pattern fails to recover and finally disintegrates.

\section{DISCUSSION}

\subsection{Geometric Cellular Automata}

Standard CAs like GoL and ECA consider only the nearest sites as neighborhood, yet more recent variants like LtL, SmoothLife and dsicrete Lenia have extended neighborhoods that enable the control over the ``granularity'' of space.  The latter ones are still technically discrete, but are approximating another class of continuous systems called Euclidean automata (EA) \cite{Pivato2007}.  We call them \textit{geometric cellular automata} (GCA).  GCAs and standard CAs are fundamentally different in a number of contrasting qualities (Table \ref{table-con}).  LtL is somehow in-between, having qualities from both sides (see Table \ref{table-ca}).

Additionally, in standard CAs, most of the interesting patterns are concentrated in specific rules like GoL, but GCAs patterns are scattered over the parameter space.  Also, the ``digital'' vs. ``analog'' distinction goes beyond a metaphor, in that many of the standard CAs are capable of ``digital'' universal computation, while whether certain kind of ``analog computing'' is possible in GCAs remains to be seen.

As GCAs being approximants of EAs, these contrasting qualities may well exist between the truly continuous EAs and discrete CAs.

\begin{table}
\centerline{\small\begin{tabular}{ccc}
\hline
Standard CA patterns & &Geometric CA patterns \\
(e.g. GoL, ECA) & &(e.g. SmoothLife, Lenia) \\
\hline
 &\textit{Structure} & \\
``Digital''& &``Analog'' \\
Non-scalable& &Scalable \\
Quantized & &Smooth \\
Localized motifs & &Geometric manifolds \\
Complex circuitry & &Complex combinatorics \\
 &\textit{Dynamics} & \\
Deterministic & &Unpredictable \\
Precise & &Fuzzy \\
Strictly periodic & &Quasi-periodic \\
Machine-like & &Life-like \\
 &\textit{Sensitivity} & \\
Fragile & &Resilient \\
Mutation sensitive & &Mutation tolerant \\
Rule-specific & &Rule-generic \\
Rule change sensitive & &Rule change adaptive \\
\hline
\end{tabular}}
\caption{Contrasting qualities in standard and geometric CAs.}
\label{table-con}
\end{table}

\subsection{Nature of Lenia}

Here we deep dive into the very nature of Lenia patterns, regarding their unpredictability, fuzziness, quasi-periodicity, resilience, and lifelikeness, at times using GoL for contrast.

\subsubsection{Persistence}
GoL patterns are either persistent, guaranteed to follow the same dynamics every time, or temporary, eventually stabilize as persistent patterns or vanish.  Lenia patterns, on the other hand, have various types of \textit{persistence}:

\begin{enumerate}[noitemsep]
\item Transient: only last for a short period.
\item Quasi-stable: able to sustain for a few to hundreds of cycles.
\item Stable: survive as long as simulations went, possibly forever.
\item Metastable: stable but transform into another pattern after slight perturbations.
\item Chaotic: ``walk a thin line'' between chaos and self-destruction.
\item Markovian: shapeshift among templates, each has its own type of persistence.
\end{enumerate}

Given a pattern, it is \textit{unpredictable} whether it belongs to which persistent type unless we put it into simulation for a considerable (potentially infinite) amount of time, a situation akin to the halting problem and the undecidability in class 4 CAs \cite{Wolfram1984}.  This uncertainty results in the vague boundaries of niches (see Figure \ref{fig-u4map}).

Even for a stable persistent pattern, in contrast to the GoL ``glider'' that will forever move diagonally, we can never be 100\% sure that an \textit{Orbium} will not eventually die out.

\subsubsection{Fuzziness}
No two patterns in Lenia are the same; there are various kinds of \textit{fuzziness} and subtle varieties.  Within a species, slightly different parameter values, rule settings, or initial configurations would result in slightly different patterns (see Figure \ref{fig-phy}).  Even during a pattern's lifetime, no two cycles are the same.

\begin{figure}
\centering
\includegraphics[width=0.7\textwidth,trim=4 4 4 4,clip]{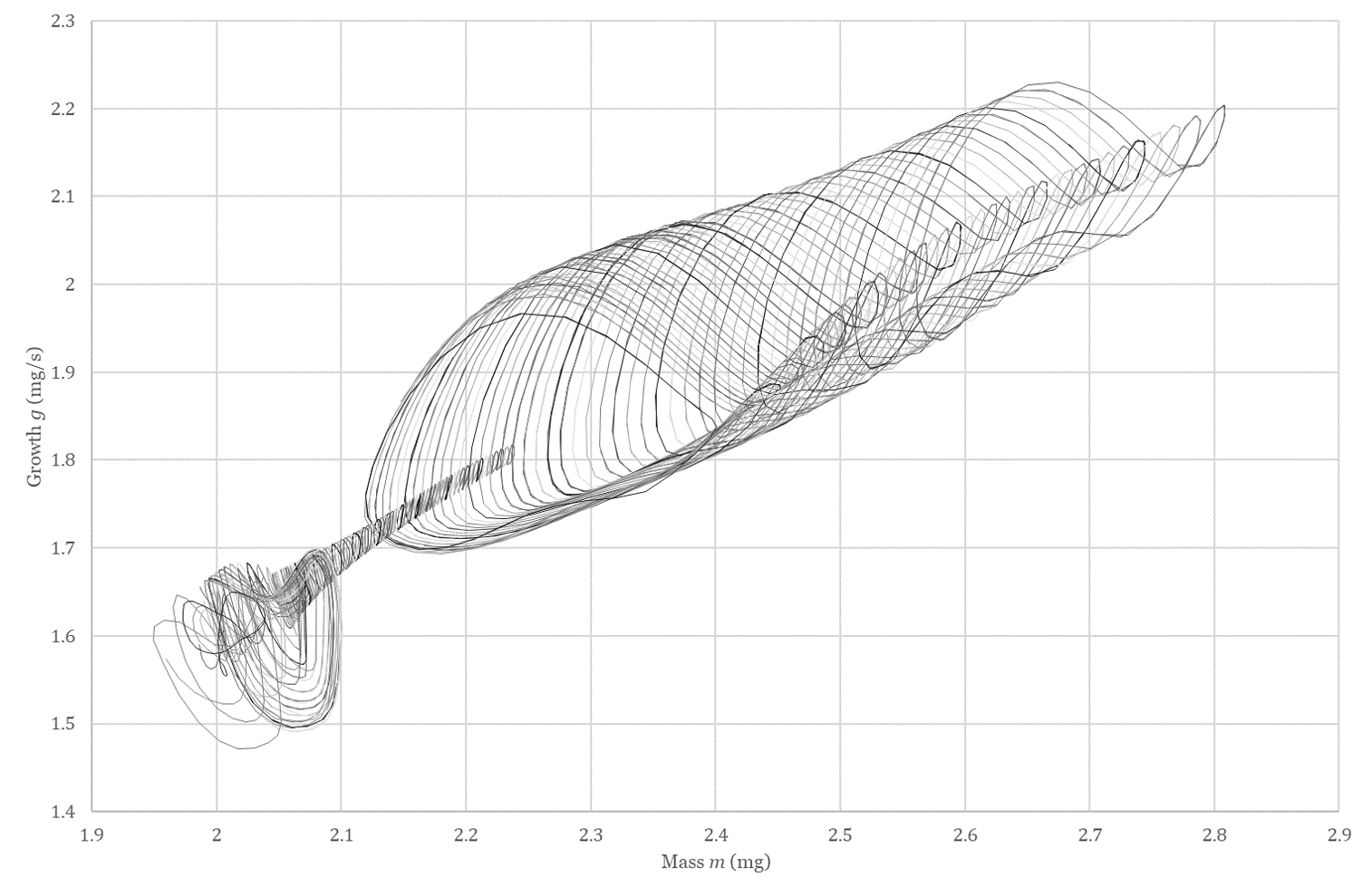}
\caption{Phase space trajectories of growth $g$ vs. mass $m$ (same cross-section as Figure \ref{fig-cross}); trajectories separated by $\Delta \sigma=0.0001$, each over a period of $t=20 \mathrm{s}$.  Each trajectory corresponds to an attractor, a group of similar trajectories hints a species-level ``attractor''.}
\label{fig-tra}
\end{figure}

Consider the phase space trajectories of recurrent patterns (Figure \ref{fig-tra}), every trajectory corresponds to an attractor (or a strange attractor if chaotic).  Yet, behind a group of similar patterns, there seems to be another kind of ``attractor'' that draws them into a common morphological-behavioral template.

\textit{Essentialism} in western philosophy proposes that every entity in the world can be identified by a set of intrinsic features or an ``essence'', be it an ideal form (Plato's idealism) or a natural kind (Aristotle's hylomorphism).  In Lenia, is there a certain kind of ``Orbium-ness'' inside all instances and occurrences of \textit{Orbium}?  Could this be identified or utilized objectively and quantitatively?

\subsubsection{Quasi-Periodicity}
Unlike GoL where a recurrent pattern returns to the exact same pattern after an exact period of time, a recurrent pattern in Lenia returns to similar patterns after slightly irregular periods or \textit{quasi-periods}, probably normally distributed.  Lenia has various types of periodicity:

\begin{enumerate}[noitemsep]
\item Aperiodic: in transient non-recurrent patterns.
\item Quasi-periodic: in quasi-stable, stable or metastable patterns.
\item Chaotic: with wide-spread quasi-period distribution.
\item Markovian: each template has its own type of periodicity.
\end{enumerate}

Principally, in discrete Lenia, there are finite, albeit astronomically large, number of possible configurations $|\mathcal{S^L}|$.  Given enough time, a recurrent pattern would eventually return to the exact same configuration (strictly periodic), an argument not unlike Nietzsche's ``eternal recurrence'', although there would be numerous approximate recurrences between two exact recurrences.  In continuous Lenia, exact recurrence may even be impossible.

\subsubsection{Plasticity}
Given the fuzziness and irregularity, Lenia patterns are surprisingly resilient and exhibit \textit{phenotypic plasticity}.  By elastically adjusting morphology and behavior, they are able to absorb deformations and transformations, adapt to environmental changes (parameters and rule settings), react to head-to-head collisions, and continue to survive.

\begin{figure}
\centering
\includegraphics[width=0.8\textwidth,trim=4 4 4 4,clip]{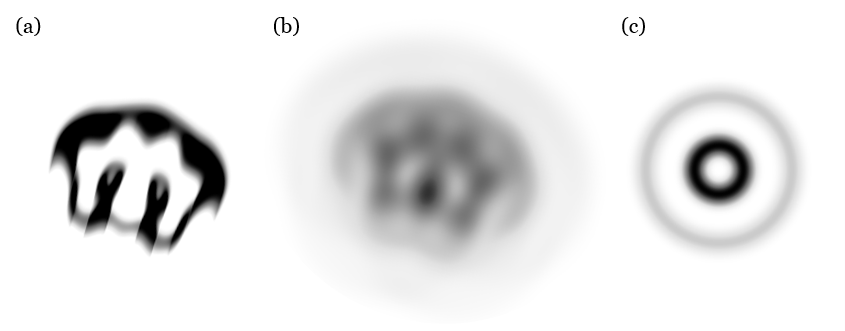}
\caption{Different views of calculation intermediates.  Configuration $\mathbf{A}^t$ (a), potential distribution $\mathbf{U}^t$ (b), and kernel $\mathbf{K}$ (c).  Notice one larger and six smaller potential peaks (b: dark spots) possibly formed by kernel resonance, and the corresponding inner spaces (a: white areas).}
\label{fig-kern}
\end{figure}

We propose a speculative mechanism for the plasticity (also self-organization and self-regulation in general) as the \textit{kernel resonance hypothesis} (Figure \ref{fig-kern}).  A network of \textit{potential peaks} can be observed in the potential distribution.  The peaks are formed by the overlapping or ``resonance'' of kernel rings casted by various mass lumps, in turn, the locations of the mass lumps are determined by the network of peaks.  In this way, the mass lumps influence each other reciprocally and self-organize into structures, providing the basis of morphogenesis.

Kernel resonance is dynamic over time, and may even be self-regulating, providing the basis of homeostasis.  Plasticity may stem from the static buffering and dynamical flexibility provided by such mass-potential-mass feedback loop.

\subsubsection{Computability}

GoL, ECA rule 110, and LtL have been demonstrated to be capable of universal computation \cite{Rendell2002, Cook2004, Evans2004}.
The proof of Turing universality of a CA requires searching for ``glider gun'' patterns that periodically emit ``gliders'' (i.e. small moving solitons), designing precise circuits orchestrated by glider collisions, and assembling them into logic gates, memory registers, eventually Turing machines \cite{Berlekamp2018}.  However, this may be difficult in Lenia due to the imprecise nature of pattern movements and collisions, and the lack of pattern-emitting configurations.

That said, particle collisions in Lenia, especially among \textit{Orbium} instances, worth further experimentation and analysis.  These have been done for classical CAs (GoL and ECA rule 100) qualitatively \cite{Martinez2017} and quantitatively using e.g. algorithmic information dynamics \cite{Zenil2018b}.

\subsection{Connections with Biological Life}

Besides the superficial resemblance, Lenia life may have deeper connections to biological life.

\subsubsection{Symmetry and Locomotion}
\begin{figure}
\centering
\includegraphics[width=\textwidth,trim=4 4 4 4,clip]{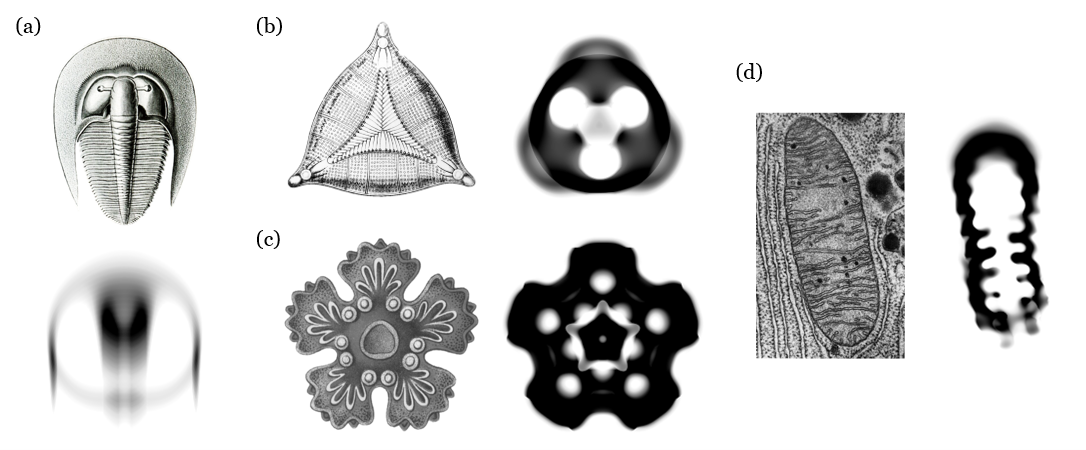}
\caption{Appearance similarities between Earth and Lenia life.  \textbf{(a)} Bilateral trilobite \textit{Bohemoharpes ungula} \cite[plate 47]{Haeckel1904} and Lenian \textit{Orbium bicaudatus}.  \textbf{(b)} Trimerous diatom \textit{Triceratium moronense} \cite[plate 4]{Haeckel1904} and Lenian \textit{Trilapillium inversus}.  \textbf{(c)} Pentamerous larva of sea star \textit{Asterias} species \cite[plate 40]{Haeckel1904} and Lenian \textit{Asterium inversus}.  \textbf{(d)} Weakly bilateral mitochondrion \cite{Porter2011} and Lenian \textit{Hydrogeminium natans}, with matrix-like internal structures.}
\label{fig-bio}
\end{figure}

Both Lenia and Earth life exhibit structural symmetry and similar symmetry-locomotion relationships (Figure \ref{fig-bio}(b-c)).

Radial symmetry is universal in Lenia order \textsf{Radiiformes}.  In biological life, radial symmetry is exhibited in microscopic protists (diatoms, radiolarians) and primitive animals historically grouped as Radiata (jellyfish, corals, comb jellies, echinoderm adults).  These radiates are sessile, floating or slow-moving, similarly, Lenia radiates are usually stationary or rotating with little linear movement.

Bilateral symmetry is the most common in Lenia.  In biological life, the group Bilateria (vertebrates, arthropods, mollusks, various ``worm'' phyla) with the same symmetry are the most successful branch of animals since the rapid diversification and proliferation near the Cambrian explosion 542 million years ago \cite{Marshall2006}.  These bilaterians are optimized for efficient locomotion, and similarly, Lenia bilaterians engage in fast linear movements.

\subsubsection{Adaptation to Environment}
The parameter space of Lenia, earlier visualized as a geographical landscape (``Ecology'' section), can also be thought of as an \textit{adaptive landscape}.  Species niches correspond to fitness peaks, indicate successful adaptation to the ranges of parameters.

Any body plan (corresponds to Earth animal phylum or Lenia family) would be considered highly adaptive if it has high biodiversity, wide ecological distribution, or great complexity.  On Earth, the champions are the insects (in terms of biodiversity), the nematodes (in terms of ecosystem breadth and individual count), and the mammals (producing intelligent species like cetaceans and primates).  In Lenia, family \textsf{Pterifera} is the most successful in class \textsf{Exokernel} in terms of diversity, niche area, and complexity.

The parallels between two systems regarding adaptability may provide insights in evolutionary biology and evolutionary computation.

\subsubsection{Species Problem}
One common difficulty encountered in the studies of Earth and Lenia life is the precise definition of a ``species'', or the \textit{species problem}.  In evolutionary biology, there exist several species concepts \cite{Mayden1997}:

\begin{itemize}[noitemsep]
\item Morphological species: based on phenotypic differentiation \cite{Darwin1859}
\item Phenetic species: based on numerical clustering (cf. phenetics) \cite{SneathSokal1973}
\item Genetic species: based on genotypic clustering \cite{Mallet1995}
\item Biological species: based on reproductive isolation \cite{Mayr1942}
\item Evolutionary species: based on phylogenetic lineage divergence \cite{Simpson1951, Hennig1966}
\item Ecological species: based on niche isolation \cite{VanValen1976}
\end{itemize}

Similar concepts are used in combination for species identification in Lenia, including morphological (similar morphology and behavior), phenetic (statistical cluster) and ecological (niche cluster) species.  However, species concepts face problems in some situations, for example, in Earth's case, species aggregates or convergent evolution, and in Lenia's case, niche complex or shapeshifting lifeforms.  It remains an open question whether clustering into species and grouping into higher taxa can be carried out objectively and systematically.

\subsection{Future Works}

\subsubsection{Open Questions}
Here are a few open questions we hope to answer:

\begin{enumerate}[noitemsep]
\item What are the enabling factors and mechanisms of how self-organization, self-regulation, self-direction, adaptability, etc. emerge in Lenia?
\item How do interesting phenomena like symmetry, alternation, metamerism, metamorphosis, particle collision, etc. arise in Lenia?
\item How is Lenia related to biological life and other forms of artificial life?
\item Can Lenia life be classified objectively and systematically?
\item Does continuous Lenia exist as the continuum limit of discrete Lenia?  If so, do corresponding ``ideal'' lifeforms exist there?
\item Is Lenia Tuning-complete and capable of universal computation?
\item Is Lenia capable of open-ended evolution that generates unlimited novelty and complexity?
\item Do self-replicating and pattern-emitting lifeforms exist in Lenia?
\item Do lifeforms exist in other variants of Lenia (e.g. 3D)? 
\end{enumerate}

To answer these questions, the following approaches of future works are suggested.

\subsubsection{More Species Data}
For the sheer joy of discovering new species, and for further understanding Lenia and artificial life, we need better capabilities in species discovery and identification.

Automatic and accurate species identification could be achieved via computer vision and pattern recognition using machine learning or deep learning techniques, e.g. training convolutional neural networks (CNNs) with patterns, or recurrent neural networks (RNNs) with time-series of measures.

Interactive evolutionary computation (IEC) currently in use for new species discovery could be advanced to allow crowdsourcing.  Web or mobile applications with intuitive interface would allow online users to simulate, mutate, select and share interesting patterns (cf. Picbreeder \cite{Secretan2008}, Ganbreeder \cite{Simon2018}).  Web performance and functionality could be improved using WebAssembly, OpenGL, TensorFlow.js, etc.

Alternatively, evolutionary computation (EC) and similar methodologies could be used for automatic, efficient exploration of the search space, as has been successfully used for evolving new body parts or body plans \cite{Kriegman2018, Jansen2008, Ha2018}.  Patterns could be represented in genetic (indirect) encoding using Compositional Pattern-Producing Network (CPPN) \cite{Stanley2007} or Bezier splines \cite{Collins2018}, which are then evolved using genetic algorithms like NeuroEvolution of Augmenting Topologies (NEAT) \cite{StanleyMiikkulainen2002}.  Novelty-driven and curiosity-driven algorithms are promising approaches \cite{LehmanStanley2011, Pugh2016, BaranesOudeyer2013}.

\subsubsection{Better Data Analysis}
Grid traversal of the parameter space (depth-first or breath-first search) is still useful in collecting statistical data, but it needs more reliable algorithms, especially for high-rank metamorphosis-prone species.

All data collected from automation or crowdsourcing would be stored in a central database for further analysis.  Using well-established techniques in related scientific disciplines, the data could be used for dynamical systems analysis (e.g. quasi-period distribution, Lyapunov exponents, transition probabilities matrix), shape analysis (computational anatomy, statistical shape analysis, algorithmic complexity \cite{Zenil2018a}), time-series analysis (cf. in astronomy \cite{Vaughan2012}), and automatic classification (unsupervised or semi-supervised learning).

\subsubsection{Variants and Generalizations}
We could also explore variants and further generalizations of Lenia, for example, higher-dimensional spaces (e.g. 3D) \cite{Bays1987, Imai2010, Hutton2012b}; different kinds of grids (e.g. hexagonal, Penrose tiling, irregular mesh) \cite{Adamatzky2006, Goucher2012, BochenekTajsZielinska2017}; different structures of kernel (e.g. non-concentric rings); other updating rules (e.g. asynchronous, heterogeneous, stochastic) \cite{Fates2013, Ryan2016, LouisNardi2018}.

\subsubsection{Artificial Life and Artificial Intelligence}
It has been demonstrated that Lenia shows a few signs of a living system:

\begin{itemize}[noitemsep]
\item Self-organization: patterns develop well-defined structures
\item Self-regulation: patterns maintain dynamical equilibria via oscillation etc.
\item Self-direction: patterns move consistently through space
\item Adaptability: patterns adapt to changes via plasticity
\item Evolvability: patterns evolve via manual operations and potentially genetic algorithms
\end{itemize}

We should seek whether these are merely superficial resemblances with biological life or are indications of deeper connections.  In the latter case, Lenia could contribute to the endeavors of artificial life in attempting to ``understand the essential general properties of living systems by synthesizing life-like behavior in software'' \cite{Bedau2003}, or could even add to the debate about the definitions of life as discussed in astrobiology and virology \cite{Benner2010, Forterre2010}.  In the former case, Lenia can still be regarded as a ``mental exercise'' on how to study a complex system using various methodologies.

Lenia could also be served as a ``machine exercise'' to provide a substrate or testbed for parallel computing, artificial life and artificial intelligence.  The heavy demand in matrix calculation and pattern recognition could act as a benchmark for machine learning and hardware acceleration; the huge search space of patterns, possibly in higher dimensions, could act as a playground for evolutionary algorithms in the quest of algorithmizing and ultimately understanding open-ended evolution. \cite{Taylor2016}

\section{Online Resources}

\begin{itemize}[noitemsep]
\item Showcase video of Lenia at \url{https://vimeo.com/277328815}
\item Source code of Lenia at \url{http://github.com/Chakazul/Lenia}
\item Source code of Primordia at \url{http://github.com/Chakazul/Primordia}
\end{itemize}

\section*{Acknowledgments}

I would like to thank David Ha, Sam Kriegman, Tim Hutton, Kyrre Glette, Pierre-Yves Oudeyer, and Lana Sinapayen for valuable discussions, suggestions and insights; Bogdan Opanchuk who actively maintains the Reikna Python library that makes GPU acceleration easy \cite{Opanchuk2018}; Mikola Lysenko who implemented FFT for SmoothLife \cite{Lysenko2012}.

\ifCS
    \bibliographystyle{ieeecs}
\else
    \bibliographystyle{ieeetr}
\fi
\bibliography{Lenia-Biology-Ref}

\end{document}